\documentclass[review,11pt]{elsarticle}
\usepackage[a4paper, margin = 1in]{geometry}
\usepackage{setspace}
\usepackage[normalem]{ulem} % For line through

\usepackage{amsmath,amsfonts}
\usepackage{algorithmic}
\usepackage{algorithm}
\usepackage{array}
\usepackage{textcomp}
\usepackage{stfloats}
\usepackage{url}
\usepackage{verbatim}
\usepackage{graphicx}
\usepackage{caption}
\usepackage{subcaption}
\usepackage{cite}
\usepackage{multirow}
\usepackage{multicol}
\usepackage{xcolor}
\usepackage{makecell}
\usepackage{mathtools}
\usepackage{enumitem}
\usepackage{xcolor}
\usepackage{float}
\usepackage{array}
\usepackage{booktabs}   % For setting \hline cell number
\usepackage{diagbox}    % Diagonal line in a table cell
\usepackage{makecell} % Allow line break in a table cell
\usepackage{hyperref}
\hypersetup{
colorlinks,
citecolor=blue,
linkcolor=blue}
\usepackage{pdflscape}  % Landscape

\usepackage{lineno}
\nolinenumbers
\usepackage{enumitem}

\newcommand*\patchAmsMathEnvironmentForLineno[1]{%
  \expandafter\let\csname old#1\expandafter\endcsname\csname #1\endcsname
  \expandafter\let\csname oldend#1\expandafter\endcsname\csname end#1\endcsname
  \renewenvironment{#1}%
     {\linenomath\csname old#1\endcsname}%
     {\csname oldend#1\endcsname\endlinenomath}}% 
\newcommand*\patchBothAmsMathEnvironmentsForLineno[1]{%
  \patchAmsMathEnvironmentForLineno{#1}%
  \patchAmsMathEnvironmentForLineno{#1*}}%
\AtBeginDocument{%
\patchBothAmsMathEnvironmentsForLineno{equation}%
\patchBothAmsMathEnvironmentsForLineno{align}%
\patchBothAmsMathEnvironmentsForLineno{flalign}%
\patchBothAmsMathEnvironmentsForLineno{alignat}%
\patchBothAmsMathEnvironmentsForLineno{gather}%
\patchBothAmsMathEnvironmentsForLineno{multline}%
}
\journal{Applied Energy}

\begin{document}

\begin{frontmatter}

\title{Graph neural networks for power grid operational risk assessment under evolving grid topology}

\author[label1]{Yadong~Zhang}
\author[label1]{Pranav~M~Karve}
\author[label1]{Sankaran~Mahadevan\corref{cor1}}
\affiliation[label1]{organization={Department of Civil and Environmental Engineering, Vanderbilt University},
            addressline={400 24th Avenue South},
            city={Nashville},
            postcode={37212},
            state={TN},
            country={USA}}
\cortext[cor1]{Corresponding author}
\ead{mahadevan.sankaran@vanderbilt.edu}

\begin{abstract}
This article investigates the ability of graph neural networks (GNNs) to identify risky conditions in a power grid over the subsequent few hours, without explicit, high-resolution information regarding future generator on/off status (grid topology) or power dispatch decisions. The GNNs are trained using supervised learning, to predict the power grid's aggregated bus-level (either zonal or system-level) or individual branch-level state under different power supply and demand conditions. The variability of the stochastic grid variables (wind/solar generation and load demand), and their statistical correlations, are rigorously considered while generating the inputs for the training data. The outputs in the training data, obtained by solving numerous mixed-integer linear programming (MILP) optimal power flow problems, correspond to system-level, zonal and transmission line-level quantities of interest (QoIs). The QoIs predicted by the GNNs are used to conduct hours-ahead, sampling-based reliability and risk assessment w.r.t. zonal and system-level (load shedding) as well as branch-level (overloading) failure events. The proposed methodology is demonstrated for three synthetic grids with sizes ranging from 118 to 2848 buses. Our results demonstrate that GNNs are capable of providing fast and accurate prediction of QoIs and can be good proxies for computationally expensive MILP algorithms. The excellent accuracy of GNN-based reliability and risk assessment suggests that GNN models can substantially improve situational awareness by quickly providing rigorous reliability and risk estimates. 

\end{abstract}

\begin{highlights}
\item Graph neural networks (GNNs) for predicting grid state over the next few hours
\item Explicit, hours-ahead operational risk assessment using GNNs
\item System, zone and branch level assessment of load shedding and branch overloading
\item Large, synthetic power grids (Case1354pegase, Case2848rte) used for demonstration
\end{highlights}


\end{frontmatter}

%%%%%%%%%%%%%%%%%%%%%%%%%%%%%%%%%%%%%%%%%%%%%%%%%%%%%%%%%%%%%%%%%%%%%%%%
%%%%%%%%%%%%%%%%%%%%%%%%%%%%% Nomenclature %%%%%%%%%%%%%%%%%%%%%%%%%%%%%
%%%%%%%%%%%%%%%%%%%%%%%%%%%%%%%%%%%%%%%%%%%%%%%%%%%%%%%%%%%%%%%%%%%%%%%%

\section*{Nomenclature}
\label{sec:nomenclature}

\singlespacing
\begin{table}[H]
    \centering
    \begin{tabular}{rl}
        $\mathcal{G}$ & Graph \\
        $\mathcal{E}$ & Set of edges in a graph \\
        $\mathcal{V}$ & Set of nodes in a graph \\
        $|\cdot|$ & Cardinality of a set \\
        $u$ & Node in a graph \\
        $\mathbf{h}^k_u$ & Embedding of $u$ in $k$-th layer \\
        $\mathbf{m}_u$ & Aggregated node embedding from neighbours of $u$ \\
        $\mathcal{N}(u)$ & Set of 1-hop neighbours of node $u$ \\
        $N_G$ & Number of generators \\
        $N$ & Number of samples\\ 
        $M$ & Number of marginal distributions \\
        $Q$ & Number of branches in a power grid \\
        $T$ & Number of time steps in the reliability unit commitment horizon\\
        $\Delta T$ & Number of time steps in the multi-step reliability and risk assessment \\ 
        $U_{G_{i,t}}$ & Generator $i$ ON (1)--OFF (0) status at $t$ \\
        $P_{G_{i,t}}$ & Power generation of generator $i$ at $t$ \\
        $R_{G_{i,t}}$ & Power generation cost of generator $i$ at $t$ \\
        $SU_{G_{i,t}}$ & Startup cost of generator $i$ at $t$ \\
        $SD_{G_{i,t}}$ & Shutdown cost of generator $i$ at $t$ \\
        $\epsilon_{\mathcal{N}}$ & Random matrix, each column follows $\mathcal{N}(0, 1)$, $\mathbb{R}^{N \times M}$ \\
        $\mathbf{x}_t$ & Random matrix, each column follows $\mathcal{N}(0, t)$, $\mathbb{R}^{N \times M}$ \\
        $\mathbf{x}^c_t$ & Random vector with covariance between columns specified by $\mathbf{C}$, $\mathbb{R}^{N \times M}$ \\
        $\mathbf{u}_t$ & CDF of $\mathbf{x}^c_t$, $[0, 1]^{N \times M}$ \\
        $\mathcal{W}_i$ & $i$-th marginal distribution \\
        $\mathbf{C}$ & Covariance matrix, $\mathbb{R}^{M \times M}$ \\
        $\mathbf{L}$ & Lower triangular matrix of Cholesky factorization of $\mathbf{C}$, $\mathbb{R}^{M \times M}$ \\
        $\Phi$ & CDF of standard normal probability distribution\\
        $\Phi_{\mathcal{W}_i}$ & CDF of the $i$-th marginal distribution\\
        $\Psi_t$ & Indicator variable $\{0,1\}$ for load shedding at time $t$ \\
        $\psi(t)$ & The amount of load shed at time $t$ (in MW) \\
        $C_s(\psi)$ & Consequence cost function for load shedding \\
        $\Gamma_t^i$ & Indicator variable $\{0,1\}$ for branch $i$ overloading at $t$ \\
        $\gamma$ & Power flow in a transmission line (network branch)\\
        $\gamma_{max}$ & Maximum allowed branch power flow \\
        $\epsilon$ & Fraction of $\gamma_{max}$ that defines the security threshold of branch power flow \\        
        $\Tilde{\gamma}$ & The amount of branch overloading, $\Tilde{\gamma} = \gamma - \epsilon \gamma_{max}$ \\
        $C_o(\Tilde{\gamma})$ & Consequence cost function for branch overloading \\
    \end{tabular}
\end{table}

\doublespacing

%%%%%%%%%%%%%%%%%%%%%%%%%%%%%%%%%%%%%%%%%%%%%%%%%%%%%%%%%%%%%%%%%%%%%%%%
%%%%%%%%%%%%%%%%%%%%%%%%%%%%% Introduction %%%%%%%%%%%%%%%%%%%%%%%%%%%%%
%%%%%%%%%%%%%%%%%%%%%%%%%%%%%%%%%%%%%%%%%%%%%%%%%%%%%%%%%%%%%%%%%%%%%%%%
\section{Introduction}
\label{sec:introduction}

The power grid is poised for a major transformation due to the increasing participation of renewable energy sources (RES), plugin devices, and flexible loads~\citep{mittelman2023potential,shang2020internet,lee2020pricing}. This paradigm shift, while pivotal in steering towards a more sustainable and diversified energy future, introduces substantial power supply and demand uncertainty ~\citep{wang2020improved,li2020review,matsuo2020investigating,ding2022distributionally} due to the volatility of RES. To effectively handle this challenge, grid operators are in need of (i) stochastic optimization techniques that take the uncertainty into account ~\citep{hlalele2020multi,ponciroli2020improved,lagos2021data,jiang2021congestion}, and (ii) rigorous methods for quantifying the operational risk (related to reserve inadequacy, load shedding, etc.). Stochastic optimization techniques are essential in grid operational decisions such as security-constrained unit commitment (SCUC)~\citep{psarros2021comparison,moretti2020efficient,isuru2020network}. On the other hand, risk quantification methods are required to understand the risks associated with the increasing uncertainty in the grid variables~\citep{sedzro2018stochastic,mohseni2022stochastic,adefarati2017reliability,stover2023reliability}. By accurately estimating hours-ahead risk, grid operators can develop strategies to ensure uninterrupted power supply at optimal costs even under volatile conditions.

Stochastic optimization methods, such as, stochastic programming~\citep{castelli2024optimal}, chance-constrained optimization~\citep{juanpera2022multi,lyu2020novel}, robust optimization~\citep{samsatli2018multi}, have been developed for performing UC under uncertainty. However, they come with their own sets of challenges. For instance, stochastic programming can only consider a small number of scenarios due to the high computational cost, resulting in limited resolution of input uncertainty representation~\citep{cao2020analyzing,rayati2022distributionally}; chance-constrained optimization seeks a solution that satisfies all probabilistic constraints, which is difficult to compute in practice~\citep{ cao2019networked}; robust optimization considers the worst possible  grid condition in decision-making process, often leading to overly conservative solution~\citep{ gu2021bridging}. Unlike UC that provide a decision for each generation unit, reliability and risk assessment in power systems has often been done at a more aggregated level, typically at the system or zone level~\citep{kirilenko2021framework,ghorani2019risk,xu2018propagating,gao2022integrated}. A range of potential contingencies, including loss of load~\citep{zhang2023power,xu2019demand,liu2022real,luo2022quantifying}, reserve inadequacy~\citep{wang2022power,li2022sizing,mohan2015efficient}, reserve inflexibility~\citep{heleno2014availability}, insufficient ramp capacity~\citep{morales2015power,gong2015ramp}, etc., have been considered in previous work. By focusing on the system or zone level, the corresponding risk assessment provides a macroscopic view of the grid’s operational state. It enables operators to identify vulnerabilities, allocate resources more effectively, and develop strategies that enhance the overall stability and safety of the power grid. The zonal/system level risk assessment requires computing the probability distribution of grid states corresponding to the given (forecast) joint probability distribution of the stochastic grid variables (wind/solar generation and load). This is accomplished by drawing Monte Carlo (MC) samples of stochastic grid variables from the forecast distribution and solving a deterministic optimization problems (usually formulated as mixed-integer linear programming (MILP)) to obtain the grid state for each MC sample. This process, however, is computationally demanding and time consuming due to the computational cost of solving numerous optimization problems. The granularity afforded by the MILP solution (bus-level grid states) is also not needed for zonal/system level risk analysis. It is therefore important to develop computational tools that can directly predict system/zone-level quantities (i.e., without solving the MILP problem) and thus enable fast and accurate reliability and risk assessment~\citep{dolanyi2022risk,ryu2020real}. Such tools need to predict the grid state corresponding to numerous future (hours-ahead) forecast scenarios of the stochastic grid variables, while accounting for the corresponding changes in the UC (grid topology) and without explicitly solving the stochastic optimization problems. They also need to be aware of and to account for the intricate interdependencies within the power grid with evolving topology, while providing accurate and rapid future grid state predictions~\citep{nikkhah2022joint,neyestani2015stochastic}. 

Fast and accurate prediction of future grid behavior under different forecast scenarios requires learning the patterns of grid behavior under different conditions. Neural networks, particularly graph neural networks (GNNs), emerge as attractive options for this task due to their superior ability to handle graph-structured information, like that encountered in power grid analysis~\citep{lee2022graph,lu2022graph,liu2020towards}. The core principle of GNN is to learn nodal features within a graph by assimilating information from neighboring nodes via message passing~\citep{feng2022powerful,liu2022introduction}. This unique capability has led to GNNs being used in various fields, including recommender system~\citep{fan2019graph}, social network analysis~\citep{min2021stgsn}, knowledge graph completion~\citep{zhang2020relational}, etc. 
In the power grid domain, GNN has been used for fault diagnosis~\citep{jacob2021fault}, power outage prediction~\citep{owerko2018predicting}, line flow control~\citep{donon2020neural}, load or generation forecasting~\citep{wu2022efficient}, and so on. In recent work~\citep{zhang2023graph,zhang2023power,falconer2022leveraging,diehl2019warm}, GNN has also been used as a surrogate for optimal power flow (OPF) computation with a fixed (known) generator on/off status (UC). However, to the best of our knowledge, the utilization of GNNs in performing prediction of system-, zone- and branch-level quantities within a planning horizon (e.g., 12 hours) under changing grid topology has not been previously reported. Such predictions are necessary for performing hours-ahead risk assessment. These predictions need to account for future operator actions corresponding to the given probabilistic forecast. For example, if the forecast shows higher than expected load demand, then the zonal/system level grid variables need to be predicted by anticipating the operator bringing more generators online to help with increasing demand. If the future operator actions and changes in the UC (grid topology) are not considered, then the risk assessment will be inaccurate.

In this work, we focus on hours-ahead prediction of quantities of interest (QoIs) at system, zone and branch levels. Accurate GNN-based system-level, zone-level or branch-level reliability and risk prediction will help operators tune their intra-day reliability unit commitments (RUCs)~\citep{chen2014applying, zhou2020multistage,park2023confidence, chen2016voltage,goleijani2013reliability, billinton2000reliability}. At the system/zone level, we consider thermal power generation and load shedding as QoIs, while transmission line flow is considered as the QoI at the branch level. Multiple GNN models are trained with each focusing on the prediction of one QoI.  Fluctuation of grid variables in one zone is often related to the changes in other zones, leading to spatial correlation. Temporal correlations also emerge due to grid inertia and sequential nature of grid operations. These correlations must be meticulously accounted for in reliability and risk assessment. For that purpose, a large number of spatio-temporally correlated samples are drawn from probability distribution forecasting of the RES power supply and load demand. These samples are then used to solve MILP problems and to obtain the grid state and QoIs corresponding to these forecast scenarios. Following the supervised learning approach, the data, i.e., samples of power supply and demand (inputs) and the corresponding QoIs (outputs), are split into training, validation and testing sets, and model evaluation is performed on the testing set. The GNN-based QoI predictions are then used for reliability and risk quantification, and the results are compared with reference solution computed using numerical (explicit stochastic optimization) approach, to evaluate the accuracy. Our main contributions include:
\begin{enumerate}[align=left, labelsep=*, leftmargin=*]
    \item Investigation of GNN models' utility to predict QoIs at system/zone/branch level under changing grid topology. This is the first investigation aimed at assessing the utility of GNNs for hours-ahead grid state prediction, which implicitly considers and accounts for the changing grid topology but does not explicitly compute future commitment/dispatch decisions. 
    \item Development of novel reliability and risk quantification methods for hours-ahead operational risk assessment, either \emph{standalone} (at a given future time instant) or \emph{multi-step} (over multiple future time instants). The developed methods can be applied at both system/zone and branch levels. This is the first attempt that considers these two important aspects of hours-ahead grid risk quantification.
    \item Development of a methodology for separating the influence of reserve-related and security-related constraints on load shedding risk quantification. Such cause-aware risk quantification can help improve the operators' situational awareness and help illuminate various risk trade-offs.
    \item Demonstration of the proposed methodology on medium to large sized synthetic power grids.  A methodology for generating spatio-temporally correlated samples of grid variables (synthetic forecasts) is developed and exercised for risk assessment of medium/large power grids. 
\end{enumerate}

The rest of the paper is organized as follows: Section~\ref{sec:background} provides a brief introduction of hours-ahead operational decision-making in a power grid, fundamentals of graph neural network, and reliability and risk assessment metrics. In Section~\ref{sec:methodology}, we discuss the elements of the proposed methodology, including GNN model development, risk quantification, and training data generation. Numerical experiments used to demonstrate the proposed methodology are described in Section~\ref{sec:numerical_experiment} and the results are summarized in Section~\ref{sec:results}. The conclusion is drawn in Section~\ref{sec:conclusion}

%%%%%%%%%%%%%%%%%%%%%%%%%%%%%%%%%%%%%%%%%%%%%%%%%%%%%%%%%%%%%%%%%%%%%%%%
%%%%%%%%%%%%%%%%%%%%%%%%%%%%% Background %%%%%%%%%%%%%%%%%%%%%%%%%%%%%%
%%%%%%%%%%%%%%%%%%%%%%%%%%%%%%%%%%%%%%%%%%%%%%%%%%%%%%%%%%%%%%%%%%%%%%%%

\section{Background}
\label{sec:background}

% \subsection{Power grid and its operation}
\subsection{Power grid operation and the SCUC problem}
\label{sec:SCUC}

The basic requirement of grid operation is to balance demand and supply, while minimizing the overall power generation cost~\citep{xavier2019transmission,vsepetanc2020convex}. The power demand fluctuates over a day; for example, power consumption increases during the morning and early evening hours, and is very low during the night. An operational strategy that aims to minimize the power generation cost requires deciding the on/off status of various generators over the operational time window (e.g., a day). Grid operators usually schedule the on/off status of generation units for a certain period (e.g., 12 hours) by employing the hours-ahead planning techniques such as security-constrained unit commitment (SCUC). The SCUC problem can be stated as:
\begin{align}
    \min_{U_{G_i}, P_{G_i}} \;\; F \left( U_{G_{i,t}}, P_{G_{i,t}} \right) = & \min_{U_{G_i}, P_{G_i}} \;\; \sum_{t=1}^{T} \sum_{i=1}^{N_G} [ U_{G_{i,t}} (1 - U_{G_{i,t-1}}) SU_{G_{i,t}} \nonumber \\
    & + U_{G_{i,t-1}} (1 - U_{G_{i,t}}) SD_{G_{i,t}} + U_{G_{i,t}} R_{G_{i,t}} (P_{G_{i,t}}) ] 
\end{align}
subject to constraints (system power balance, power generation limits, minimum on/off time constraints, ramping rate constraints, reserve requirements, line flow limits, etc.). We refer the reader to~\citep{yang2021comprehensive} for additional details of the SCUC problem.

Given the forecast of demand and supply, the outcome of SCUC plays a fundamental role in subsequent grid operation. For instance, due to its strategically important position in the operational framework, SCUC directly influences the calculation of economic dispatch, which aims at refining the generation schedule every five to ten minutes~\citep{lyu2020novel,tang2018novel}. A rigorous hours-ahead or intra-day grid risk assessment methodology needs to consider the evolving on/off status of the generators in response to the changing grid state. When sampling-based operational risk assessment methods are used, this necessitates solving thousands of SCUC problems corresponding to the forecast scenarios. Although various advanced algorithms (branch-and-bound~\citep{shafie2011unified}, Lagrangian relaxation~\citep{jiang2012parallel}, Benders decomposition~\citep{laothumyingyong2010security}, etc.) have been developed for solving the SCUC problem, this task is still computationally prohibitive, especially for large power grids with thousands of buses and branches. 

To address this challenge, data-driven surrogate models are being developed in the literature. These models can be categorized into two types: \textit{end-to-end} and \textit{assistant} models.
End-to-end models, including decision tree~\citep{yang2021machine}, gated recurrent neural network~\citep{yang2022deep}, etc.,  are trained to generate direct predictions of UC or ED given demand and RES power generation as input. These models are capable of predicting thousands of predictions of QoIs corresponding to various demand/supply scenarios in a short time (typically in milliseconds). However, the presence of both discrete and continuous variables poses considerable challenge to the accuracy these models. Data-driven approaches for solving MILP is still an active research topic. Moreover, most of the existing end-to-end models can only predict 2 to 5 hours ahead~\citep{ramesh2023feasibility,ramesh2023spatio}, while intra-day or day-ahead grid operation usually requires a longer scheduling horizon, e.g., 12 hours. These limit the utilization of currently available end-to-end models. In contrast, the \emph{assistant} models speed up a particular subtask in the SCUC solution process, such as identifying the active constraints~\citep{yang2021machine} (critical branches that probably exceed line flow limit~\citep{mohammadi2021machine}, generators that may approach the maximum generation capacity~\citep{hasan2022topology}, etc.), or providing a better initial guess (warm start)~\citep{ramesh2022machine,schmitt2022fast}, given the (forecasting) distribution of demand and RES power supply. The active constraints are retained during the SCUC problem solution process, whereas inactive constraints are removed to decrease the computational cost. A warm start also helps reduce the computational cost of the SCUC solution. Nevertheless, the speed up is limited as the iterative SCUC problem solution is still needed when using an assistant model.

In real-world grid operation, operators are often interested in risk quantification of the entire power grid or specific zones~\citep{kirilenko2021framework,xu2018propagating,gao2022integrated}. For this purpose, a machine learning model that can directly predict zonal/system level quantities, without requiring explicit SCUC problem solutions, could be very useful. This approach has not been investigated and/or reported in the literature. The objective of this paper is to investigate the performance of graph neural networks for such, direct zonal/system level risk estimation.

\subsection{Graph neural network (GNN)}

Graph neural network (GNN) is a type of neural network that can process graph-structured data~\citep{zhou2020graph,wu2020comprehensive,liu2020towards}. The core of GNN is message passing mechanism, which systematically aggregates and updates node features by assimilating information from the neighboring nodes~\citep{feng2022powerful,liu2022introduction}. Within the topology of a graph, nodes directly linked to a given node are designated as its \emph{one-hop} neighbors. Extending this nomenclature, the immediate neighbors of these one-hop nodes are identified as \textit{two-hop} neighbors, and so on. Each layer of a GNN model is dedicated to orchestrating the exchange of information between a node and its one-hop neighbors, and multiple layers can be stacked together to assimilate information from remote neighbors in the graph. Message passing is comprised of \textit{aggregation} and \textit{updating}, which can be represented as:
\begin{align}
    \mathbf{m}_u &= \text{\textbf{AGGREGATE}}(\mathbf{h}^k_v), \; v \in \mathcal{N}(u), \\
    \mathbf{h}_u^{(k+1)} &= \text{\textbf{UPDATE}}(\mathbf{h}_u, \; \mathbf{m}_u).
\end{align}

GNNs have been remarkably successful in recommender systems~\citep{fan2019graph}, knowledge graph enhancements~\citep{zhang2020relational}, social network analyses (anomaly detection and community identification)~\citep{min2021stgsn}, etc. One of the earliest GNN benchmarks utilizes a graph convolutional network (GCN), which considers symmetric-normalized aggregation with self-loop~\citep{kipf2016semi}. GNNs have been used in power grid applications for fault detection~\citep{sapountzoglou2020}, outage prediction~\citep{benmahamed2017}, etc. However, GCN needs to access the entire graph in every aggregation and update operation, which is computationally expensive and hinders its use for large graphs. To alleviate the computational burden, another category of GNNs that can work on partial graphs has been developed. GraphSAGE~\citep{hamilton2017inductive} is one such GNN. It generates embeddings by sampling a fixed-size neighborhood and then aggregating the features from these neighbors. GraphSAGE can be represented as:
\begin{align}
    \mathbf{m}_u = \text{CONCAT} \left( \mathbf{h}_u, \;\; \frac{1}{|\mathcal{N}(u)|} \sum_{v \in \mathcal{N}(u)} \mathbf{h}_v \right).
\end{align}
GraphSAGE does not require the entire graph to be loaded into the computer memory. It gradually samples nodes over the graph, making it attractive for the learning task on large graphs. In this work, we explore GraphSAGE's utility for learning the grid behavior pattern over a few hours into the future, for large grids (thousands of buses).

\subsection{Reliability and risk in the power grid operation -- methods and metrics}
\label{sec:reliability_risk_matrics}

In recent years, there has been a push to shift power grid operational decision-making from experience- or judgement-based best practices (e.g., minimum reserve requirement) to advanced algorithms (e.g., stochastic dynamic reserve allocation~\citep{cao2020analyzing}, stochastic SCUC/SCED~\citep{rayati2022distributionally}, and robust unit commitment~\citep{cao2019networked}, etc.) that are able to consider variability in supply and demand. These algorithms have allowed decision-making under increasing supply/demand volatility due to the increasing penetration of renewable energy. However, they do not  explicitly quantify or communicate the risk associated with UC/ED portfolio selection, and therefore do not support systematic cost versus risk trade-off analysis. Such analyses could be very important tools for grid operators. Furthermore, many of the advanced decision-making methods treat power grids as purely financial systems instead of physical ones. Therefore, it is not possible to obtain quantification of the physical asset risk and the financial risk (monetary loss) using the aforementioned algorithms.

In a recent work~\citep{stover2023reliability}, Stover et al. provide a comprehensive review of state-of-the-art reliability and risk quantification methods, and propose a new sampling-based framework that encompasses various failure modes. This framework develops metrics at three levels to characterize the risk: the conditional expectation of the failure event (Level 1), the probability of this event occurring (Level 2), and the consequence (monetary cost) (Level 3). In this study, we place emphasis on the Level 2 and Level 3 metrics. The Level 2 metric captures the probability of failure with respect to a failure mode, e.g., reserve inadequacy, loss of load. The Level 3 metric adopts a more holistic view, not only gauging the likelihood of failure but also factoring in the economic implications or consequence costs linked to the adverse event. This dual consideration ensures a comprehensive understanding of the operational risk, melding both the probabilistic and economic dimensions of potential system failures. 

However, the above framework only considers reliability and risk metrics at system level, while risk in individual branches is not quantified. In the context of increasing utilization of renewable energy, the volatility of transmission line flow is also significantly elevated, introducing additional risk in grid operation. As a  result, it is also necessary to take this kind of risk into consideration. In this work, we utilize the risk framework in \citep{stover2023reliability} for hours-ahead zonal/system level reliability and risk prediction under \emph{temporally evolving} grid topology (i.e., unit commitments), and extend the analysis to branch-level QoIs.

%%%%%%%%%%%%%%%%%%%%%%%%%%%%%%%%%%%%%%%%%%%%%%%%%%%%%%%%%%%%%%%%%%%%%%%%
%%%%%%%%%%%%%%%%%%%%%%%%%%%%% Methodology %%%%%%%%%%%%%%%%%%%%%%%%%%%%%%
%%%%%%%%%%%%%%%%%%%%%%%%%%%%%%%%%%%%%%%%%%%%%%%%%%%%%%%%%%%%%%%%%%%%%%%%
\section{Proposed Methodology}
\label{sec:methodology}

Three key elements of the proposed methodology, namely, GNN model development, reliability and risk assessment, and training data generation, are discussed in this section.

\subsection{GNN model development}
\label{sec:model_development}

\begin{figure}
    \centering
    \includegraphics[width=0.8\linewidth]{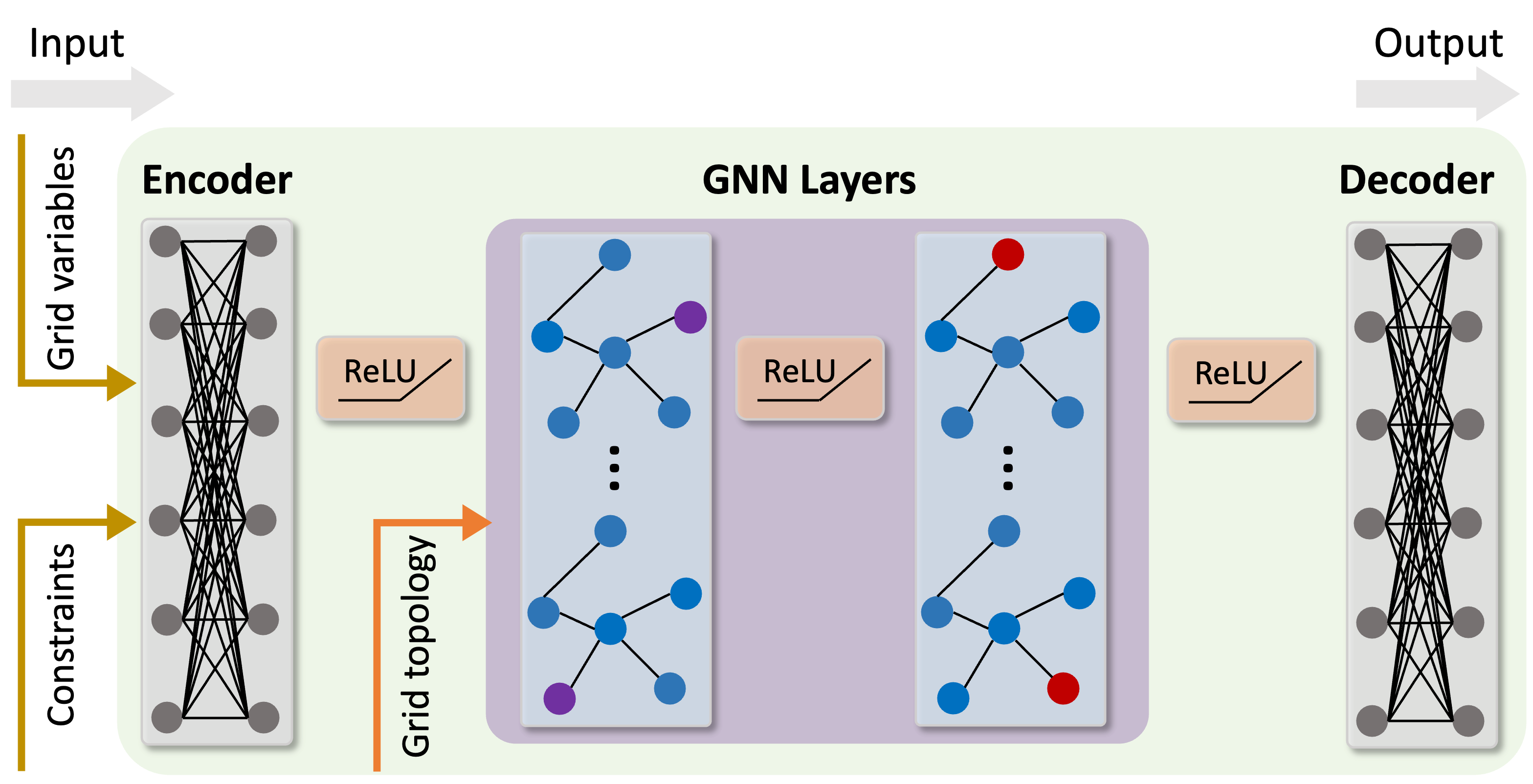}
    \caption{Schematic of the GNN surrogate model.}
    \label{fig:GNN_illustration}
\end{figure}

In this work, we utilize GNN models consisting of three components: \textit{Encoder}, \textit{GNN layers}, and \textit{Decoder}. \textit{Encoder} generates a high-dimensional representation (\textit{embedding}) of the input vector, i.e., node features. This embedding acts as an enriched representation of the input data, capturing intricate patterns and relationships inherent in the input. Once the encoder processes the input, it is passed on to the \textit{GNN layers}, which aggregate and update node representations by incorporating information from neighboring nodes, allowing the network to understand the broader structure and context of the graph. Such aggregation aids in refining the node embedding by assimilating both local and global information from the graph. Finally, the \textit{Decoder} takes these enhanced node representations and translates them into the desired output format. Depending on the specific application, the decoder might reconstruct the original data, predict certain properties of the graph, or even generate new graph structures. The combination of \textit{Encoder}, \textit{GNN layers}, and \textit{Decoder} ensures that the GNN model can efficiently capture, process, and utilize graph-structured data. Note that both \textit{Encoder} and \textit{Decoder} consist of two fully connected neural layers (multi-layer perceptron, or MLP). Park et al.~\citep{park2023confidence} have shown that the expressive power of GNN models can be improved by adding these fully connected layers. The architecture of the GNN surrogate model is shown in Fig.~\ref{fig:GNN_illustration}.

Three separate GNN models are constructed for predicting different QoIs: (\textit{i}) aggregated thermal power generation, (\textit{ii}) aggregated load shedding, and (\textit{iii}) branch power flow. The first two models perform graph-level prediction and output a single value at each time step; therefore, a graph pooling layer is added before the \textit{Decoder} to generate global embedding. The third model (model \textit{iii}) performs node-level prediction, i.e., net power generation at individual bus (+ for generation and - for demand), which can be converted to branch flow $\gamma$ using the power transfer distribution factor (PTDF) matrix. A ReLU activation function is added after each neural network layer except for the last one. These GNN models take grid topology, load demand and RES power generation as inputs, and generate predictions of the QoIs. Following a supervised learning approach, numerical solutions of QoIs (from SCUC analysis) are used as outputs (ground truth) in the training process. The loss function consists of mean squared error (MSE) and a regularization term to ensure that predictions are within reasonable upper and lower bounds.

\subsection{Reliability and risk assessment}
\label{sec:reliability_risk_assessment}

We consider two failure modes for reliability and risk assessment: \textit{load shedding} and \textit{branch overloading}. Load shedding refers to the deliberate reduction or shutdown of electrical power supply to a consumer in a system to prevent the failure of the entire system, when demand strains the capacity of the system. Branch overloading implies that the electric current on a branch exceeds a safety threshold and may lead to failure. We perform two types of hours-ahead reliability and risk assessment: (a) \emph{standalone} hours-ahead assessment that considers QoIs at a specific future time instant $(t + i)$, and (b) \emph{multi-step} hours-ahead assessment that considers QoIs at multiple time instants ($(t + i), (t + i) + 1, (t + i) + 2, \dots$) in order to provide an aggregated estimate over a future duration. We use the variable $\Delta T$ to distinguish the reliability and risk aggregation horizon from the RUC planning horizon $T$. 

\subsubsection{Load shedding} 

Both standalone and multi-step load shedding are considered in the load shedding analysis. Standalone shedding represents the immediate vulnerability within the system, highlighting areas where current demand exceeds the system's capacity or where there are breakdowns at a particular time instant. On the other hand, the multi-step shedding is a forward-looking measure that identifies potential pitfalls of the grid operation over multiple future time instants in the near future. This dual perspective establishes a comprehensive assessment of the system’s reliability that can account for both current and future challenges. The probabilities of standalone and multi-step load shedding are defined as:

\noindent \textit{standalone}:
\begin{align}
\mathcal{P}_s(t) &= p(\Psi_t = 1),
\end{align}

\noindent \textit{multi-step}:
\begin{align}
\mathcal{P}_s^+(t) &= p\left(\bigcup_{t'=t+1}^{t + \Delta T} \Psi_{t'}=1 \, \vert \, \Psi_t=1 \right),
\end{align}
respectively.

Load shedding usually leads to compromised service reliability, potential economic losses, and safety concerns, hence there is consequence (monetary cost) associated with it. In~\citep{stover2023reliability}, the degree of shedding is considered and the corresponding cost is defined as:
\begin{align}
    \label{eqn:shedding_unit_cost}
    R_s(t) = \int^{\psi(t)}_0 C_s(\psi) \: d\psi,
\end{align}
Note that $C_s(\psi)$ depends solely on $\psi$, while $\psi(t)$ is a function of $t$. The associated risk is then quantified as the expectation of potential monetary cost:
\begin{align}
\mathcal{R}_s(t) = \mathbb{E}\left[ R_s(t) \right],
\end{align}
and
\begin{align}
\mathcal{R}_s^+(t) = \sum_{\Delta t = 1}^{\Delta T} \mathbb{E}\left[ R_s(t + \Delta t) \right],
\end{align}
for standalone and multi-step load shedding, respectively. Here, we assume that the cost function $C_s(\psi)$ is temporally invariant. However, the consequence cost for load shedding events further away from the current time could be reduced by considering suitable, temporally diminishing cost functions, e.g., by using a discount factor $1/(1+\Delta t)$ to modify $C_s(\psi)$. The overall risk at $t$ can be computed as:
\begin{align}
\Re_s(t) &= \mathcal{R}_s(t) + \mathcal{R}_s^+(t) \nonumber \\
&= \mathbb{E}\left[ R_s(t) \right] + \sum_{\Delta t = 1}^{\Delta T} \mathbb{E}\left[ R_s(t + \Delta t) \right] \\
&= \sum_{\Delta t = 0}^{\Delta T} \mathbb{E}\left[ R_s(t + \Delta t) \right] \nonumber
\end{align}

\subsubsection{Branch overloading}

Branch overloading can result in a large amount of heat generation that may weaken the insulating material, accelerate the aging process of the conductor, and even lead to equipment failure. Hence it is vital to monitor and manage branch loading status. Once again, we consider both standalone and multi-step risk. The respective probabilities are calculated as:

\noindent \textit{standalone}:
\begin{align}
\mathcal{P}^I_o(t) &= p\left( \Gamma_t=1 \right),
\end{align}

\noindent \textit{multi-step}:
\begin{align}
\mathcal{P}^{I+}_o(t) &= p\left( \bigcup_{t'=t+1}^{t+\Delta T} \Gamma_{t'}=1 \, \vert \, \Gamma_t=1 \right),
\end{align}

In real-world grid operations, electric current through any given branch is influenced by the flows in other branches. This interdependence is due to the physical laws that govern electrical flow (Kirchhoff’s laws). Therefore, we compute conditional probabilities (of one branch overloading given overloading in other branches) for both standalone and multi-step cases. The corresponding probabilities are computed as:

\noindent \textit{standalone (conditional)}:
\begin{align}
\left. \mathcal{P}^{II}_o(t) \right\vert_{i,j} &= \left[ p\left( \Gamma_t^j=1 \, \vert \, \Gamma_t^i=1 \right) \right]_{i, j},
\end{align}

\noindent \textit{multi-step (conditional)}:
\begin{align} \left. \mathcal{P}^{II+}_o(t) \right\vert_{t',i,j} &= \left[ p\left(\bigcup_{t'=t+1}^{t+\Delta T} \Gamma_{t'}^j=1 \, \vert \, \Gamma_t^i=1 \right) \right]_{t',i, j}.
\end{align}

The consequence of branch overloading, also with consideration of the degree of overloading, is defined as:
\begin{align} R_o(t) = \max \biggl\{ \int_0^{\Tilde{\gamma}(t)} 
    \label{eqn:overloading_unit_cost}
    C_o(\Tilde{\gamma}) \: d \Tilde{\gamma}, \;\;\; 0 \biggl\}.
\end{align}
The risk of standalone and multi-step load shedding can be computed as:
\begin{align} 
    \mathcal{R}_o^I(t) &= \sum_{i=1}^Q\mathbb{E} \left[ R_o(t) \right]_i, \\ \mathcal{R}_o^{I+}(t) &= \sum_{\Delta t = 1}^{\Delta T} \sum_{i=1}^Q\mathbb{E} \left[ R_o(t + \Delta t) \right]_i,
\end{align}
and the overall risk can be obtained as:
\begin{align} 
    \Re_o(t) &= \mathcal{R}_o^I(t) + \mathcal{R}_o^{I+}(t) \nonumber \\ &= \sum_{i=1}^Q\mathbb{E} \left[ R_o(t) \right]_i + \sum_{\Delta t = 1}^{\Delta T} \sum_{i=1}^Q\mathbb{E} \left[ R_o(t + \Delta t) \right]_i \\ &= \sum_{\Delta t = 0}^{\Delta T} \sum_{i=1}^Q\mathbb{E} \left[ R_o(t + \Delta t) \right]_i \nonumber
\end{align}

Note that the branch overloading risk formula considers branch overloading risk for all transmission lines for the current (standalone) and future (multi-step) time horizons. The interaction between different branches is included in this computation. 

\subsection{Training data generation} 
\label{sec:training_data_generation}

Demonstration of the proposed methodology requires probabilistic forecasts for the grid. Since such forecasts are typically not available for synthetic power grids, we develop a method here for generating synthetic forecasts. We utilize an autoregressive model to account for temporal correlations for a given grid variable over the future time steps of interest. The spatial dependency between different grid variables is considered by specifying a covariance matrix. By incorporating both temporal and spatial correlation, our methodology offers a robust framework for a nuanced simulation of grid dynamics, which is valuable in demonstrating the utility of probabilistic grid behavior analysis methods. Given a planning horizon (time steps) $T$ and probabilistic distributions $\mathcal{W}_i$ ($i \in {1,2,\dots M}$) of grid variables, the procedure for generating $N$ spatio-temporally correlated training data samples for all time steps is shown in Algorithm~\ref{alg:alg1}.
\begin{algorithm}[!ht]
\caption{Sampling spatio-temporally correlated stochastic grid variables}\label{alg:alg1}
\begin{algorithmic}

\STATE \textbf{Inputs}: $\#$samples $N$, $\#$time steps $T$, marginals $\mathcal{W}_1$, $\mathcal{W}_2$, ..., $\mathcal{W}_M$, covariance matrix $\mathbf{C} \in \mathbb{R}^{M \times M}$

\STATE \textbf{Start} 

\STATE \textbf{do}: $\mathbf{L} \gets  \mathbf{C} = \mathbf{L}\mathbf{L}^T$, $\mathbf{x}_0 \gets \mathbf{0} \in \mathbb{R}^{N \times M}$

\STATE \textbf{for} t = 1:T \textbf{do}:

\STATE \hspace{0.5cm} $\mathbf{x}_t \gets \mathbf{x}_{t-1} + \epsilon_{\mathcal{N}}, \; \epsilon_{\mathcal{N}} \sim \mathcal{N}(0, 1)$ \hspace{5em} % $\triangleright$ Random walk Markov chain

\STATE \hspace{0.5cm} $\mathbf{s}_t \gets \mathbf{x}_t / \sqrt{t}$ \hspace{5em} % $\triangleright$ Convert to correlated standard normal

\STATE \hspace{0.5cm} $\mathbf{x}^c_t \gets \left( \mathbf{L} \mathbf{s}'_t \right)'$ \hspace{5em} % $\triangleright$ Get CDF of correlated standard normal variables

\STATE \hspace{0.5cm} $\mathbf{u}_t \gets \Phi\left(\mathbf{x}^c_t\right)$ \hspace{5em} % $\triangleright$ Get CDF of correlated standard normal variables

\STATE \textbf{End for}

\STATE \textbf{for} t = 1:$T$ \textbf{do}:

\STATE \hspace{0.5cm} \textbf{for} i = 1:M \textbf{do}:

\STATE \hspace{1cm} $\mathbf{w}^i_t \gets \Phi_{\mathcal{W}_i}^{-1} \left( \mathbf{u}_t^i \right)$, $\mathbf{u}_t^i \in \mathbb{R}^N $ \hspace{1em} % $\triangleright$ Apply inverse CDF to get correlated random variables from specified marginals

\STATE \textbf{End for}

\STATE \textbf{End}

\STATE \textbf{Outputs}: $ \mathbf{W}_t = [\mathbf{w}^1_t \; \mathbf{w}^2_t \; \dots \; \mathbf{w}^M_t]$ consisting of stochastic variables with correlations specified by $\mathbf{C}$. (The $N$ variables in the same column are from the same marginal probability distribution).

\end{algorithmic}
\end{algorithm}

\begin{figure}[H] 
    \centering \includegraphics[width=0.8\linewidth]{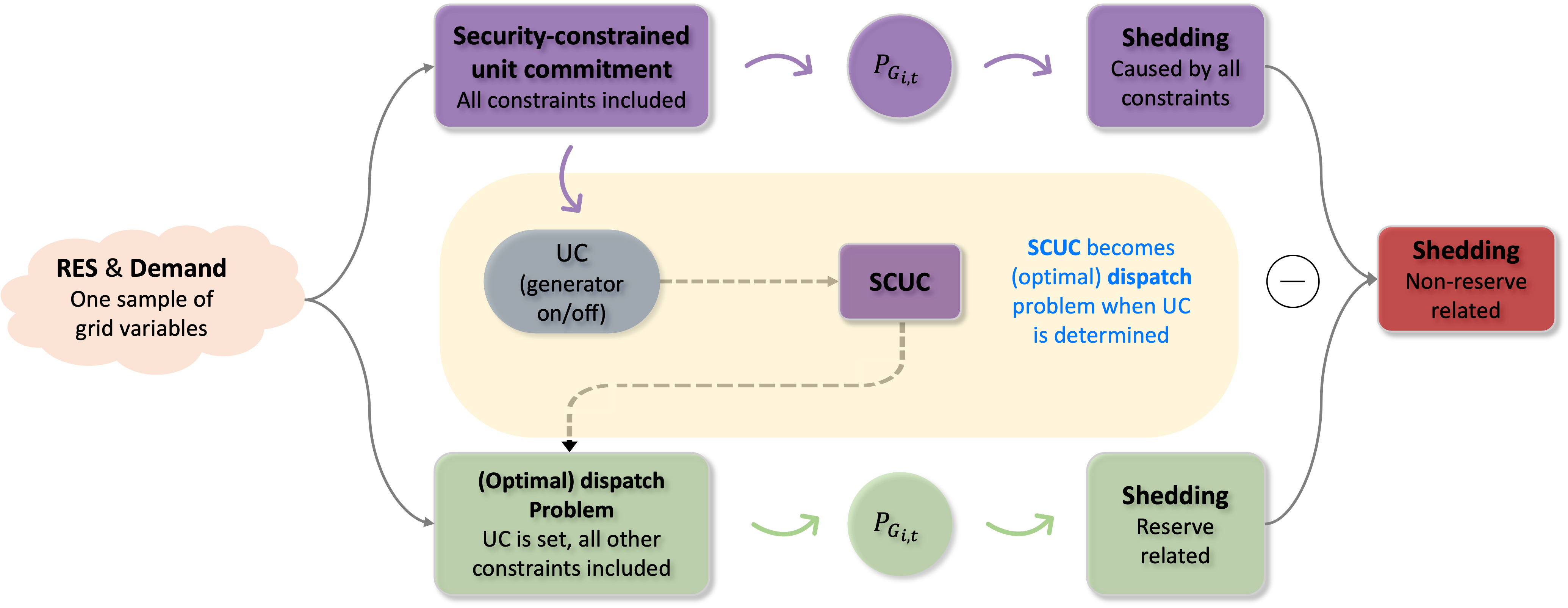} 
    \caption{Separating the influence of reserve and non-reserve constraints on load shedding. SCUC is solved twice: 1) \textcolor{violet}{violet labeled} where all constraints are considered,  the decision variables include UC and generator dispatch  and suggested load shedding is due to reserve and non-reserve constraints; 2) \textcolor{green}{green labeled} where previous UC solution is used (i.e., the reserve capacity is pre-determined), generator dispatch is the only decision variable and load shedding depends on reserve related constraints. The non-reserve load shedding is thus obtained by subtracting the latter from the former.}
    \label{fig:pipeline}
\end{figure}

We also develop a methodology to generate load shedding cause-aware training data to train GNN models that could identify the cause of (future) load shedding. There are two main causes of load shedding: reserve shortage and security constraint (i.e., transmission line flow limit) violations. Reserve shortage occurs when backup power is not available to meet the demand in the event of a sudden drop in supply. Reserve shortage could occur due to lack of adequate power generation, inaccurate demand forecasting, or disruptions in fuel supply. If the set of online generators cannot provide the minimum required reserve supply, the SCUC optimizer  may shed load to ensure that sufficient reserves are available. In other words, the amount of load shedding caused by potential reserve shortage depends on generator on/off condition, and such \textit{reserve related} shedding is, to some extent, determined once UC is given. Security constraints, on the other hand, refer to the physical limits of power transmission and distribution infrastructure. These constraints could be violated due to outdated equipment, maintenance issues, or bottlenecks in certain parts of the grid. Even if an adequate amount of power is being generated, it cannot be effectively distributed to all areas if the security constraints are being violated. This leads to targeted load shedding to prevent equipment overload. When predicting future load shedding, it is of interest to the operator to know the cause of the problem (either insufficient reserves or security constraint violations). This will allow the operator to implement a suitable problem-specific mitigation strategy. We propose a methodology to identify the cause of load shedding in a SCUC solution of the optimal power flow problem, while generating data to train the GNN models. Specifically, we propose to compute dispatch under two sets of constraints: (a) all  constraints (reserve shortage and security constraints), and (b) only non-reserve constraints. We record the load shedding proposed by the numerical solver in these two cases to separately quantify reserve-related and non-reserve related load shedding for a given grid forecast, as shown in Fig.~\ref{fig:pipeline}. 

Theoretically, this simple separation is a bit rough as there can also be load shedding caused by security constraints even if UC is given (corresponding to computing the (optimal) dispatch). However, it can at least provide an intuition of the relative contribution of reserve and non-reserve factors to load shedding. With such intuition, operators can take targeted actions to better maintain supply/demand balance with less effort and/or economic cost.

%%%%%%%%%%%%%%%%%%%%%%%%%%%%%%%%%%%%%%%%%%%%%%%%%%%%%%%%%%%%%
%%%%%%%%%%%%%%%%%%%%%%%%%%%%%%%%%%%%%%%%%%%%%%%%%%%%%%%%%%%%%
%%%%%%%%%%%%%%%%% Numerical Experiments %%%%%%%%%%%%%%%%%%%%%
%%%%%%%%%%%%%%%%%%%%%%%%%%%%%%%%%%%%%%%%%%%%%%%%%%%%%%%%%%%%%
%%%%%%%%%%%%%%%%%%%%%%%%%%%%%%%%%%%%%%%%%%%%%%%%%%%%%%%%%%%%%

\section{Numerical Experiments}
\label{sec:numerical_experiment}

We illustrate the utility of the GNN models using numerical experiments conducted on three synthetic grids: Case118~\citep{zimmerman1997matpower}, Case1354pegase~\citep{fliscounakis2013contingency} and Case2848rte~\citep{josz2016ac}. About 20\% generators in each of the synthetic grids are taken to be RES generation units. The grids are partitioned into three, eight and sixteen zones, respectively, and the implementation details can be found in~\ref{sec:power_grids}. We highlight the following:
\begin{itemize}[align=left, labelsep=*, leftmargin=*] 
    \item Grids are partitioned based on the geographic locations of buses. The details of zonal partitions can be found in~\ref{sec:power_grids}.  
    \item Wind turbines are used to represent RES generators, and wind power will be preferably used to meet the load demand before thermal generators are taken into consideration. 
    \item Wind power is obtained by considering the probability distribution of  wind speed. A given realization of wind speed is converted to wind power using an assumed power rating curve.  
    \item Without loss of generality, the maximum generation capacity of all wind turbines is considered to be the same. The maximum generation capacity for thermal generators is randomly sampled from uniform distribution. 
    \item Aggregated load demand and wind power generation are used for each zone. These aggregated values are utilized in modeling the spatio-temporal dependence between different zones, while a fixed scalar distribution factor is used to distribute these aggregated values to individual buses in the same zone~\citep{zhang2023graph}.
\end{itemize}

We consider an intra-day planning horizon of $T = 12$ hours for SCUC calculation. Stochastic grid variables are sampled using Algorithm~\ref{alg:alg1}. Specifically, wind speed is assumed to follow a Weibull probability distribution (and converted to wind power) while load is described by a truncated normal probability distribution. The Latin hypercube sampling (LHS) technique is utilized for sampling the grid variables in the first time step. The samples are used as inputs to an MILP solver to obtain the power grid state under the forecast distributions of the grid variables. The well-known power system optimal scheduling tool MATPOWER~\citep{zimmerman1997matpower} is used to formulate the MILP problems, which are then solved by using the state-of-the-art numerical solver Gurobi~\citep{gurobi2021gurobi}. Overall, 1000 SCUC  solutions us are generated for model training and evaluation. The loss \textit{vs} number of training data suggests that it is not necessary to use more samples in training process, and similar observation has been reported in~\citep{zhang2023graph,ramesh2023spatio}.

The GNN models are trained using a supervised learning approach in Python routines  PyTorch and PyG.  Inputs to the model include wind generation and load, the grid topology and branch properties. Outputs are the corresponding system/zonal aggregated thermal power generation and load shedding, as well as branch flow, for the planning period (12 hours). The 1000 MILP-based SCUC solution samples are split into training, validation, and testing sets with proportions 70\%, 10\%, 20\% respectively. Note that it is not necessary to use more training data as suggested by~\citep{zhang2023power}. Details of GNN model architecture can be found in~\ref{sec:GNN_details}. The GNN models are trained on a computer with 128 GB Intel Core i9-13900K processor and  RTX A6000 graphical card with 48 GB RAM.

For the multi-step reliability and risk assessment, we consider $\Delta T = 2$ hours. This means that we compute aggregated reliability and risk over two hour-long windows starting at multiple future time instants. For demonstration purposes, we utilize a constant cost function by setting $C_s = \$10/MW$ and $C_o = \$1/MW$ in Eq.~\ref{eqn:shedding_unit_cost} and \ref{eqn:overloading_unit_cost}, respectively. Note, however, that it is possible to implement any cost function. We conduct experiments on the system and individual zones simultaneously, while the results are evaluated separately. By doing so, we aim at investigating the relative vulnerability of different zones and their contributions to the entire power grid. We identify important branches in each grid by computing the average of maximum power flow in all the branches over all the training data samples. We refer to these branches  as \textit{significant} branches. We  focus on these branches for branch overloading reliability and risk assessment, since the loading percentage on the remaining branches is far below the security threshold in $>90$\% operating conditions. The fraction of branch power flow security threshold is set as $\epsilon = 85\%$.

%%%%%%%%%%%%%%%%%%%%%%%%%%%%%%%%%%%%%%%%%%%%%%%%%%%%%%%%%%%%%%%%%%%%%
%%%%%%%%%%%%%%%%%%%%%%%%%%%% Results %%%%%%%%%%%%%%%%%%%%%%%%%%%%%%%%
%%%%%%%%%%%%%%%%%%%%%%%%%%%%%%%%%%%%%%%%%%%%%%%%%%%%%%%%%%%%%%%%%%%%%

\section{Results}
\label{sec:results}

The results of the numerical experiments with the three grids are detailed in this section. Firstly, to show that the grid topology indeed changes over the 12-hour planning period, we plot the rate of change of generator ON/OFF status for all three grids.  We then evaluate the performance of the GNN models in predicting the QoIs, and the results are shown in terms of mean relative error (MRE, \%). Finally, we evaluate the accuracy of reliability and risk quantification based on the GNN predictions by comparing them with the reference reliability and risk estimates computed based on the MILP solution.

\subsection{Evolution of thermal generator on/off status and grid topology}
\begin{figure}[!ht]
    \centering
    \includegraphics[width=0.5\linewidth]{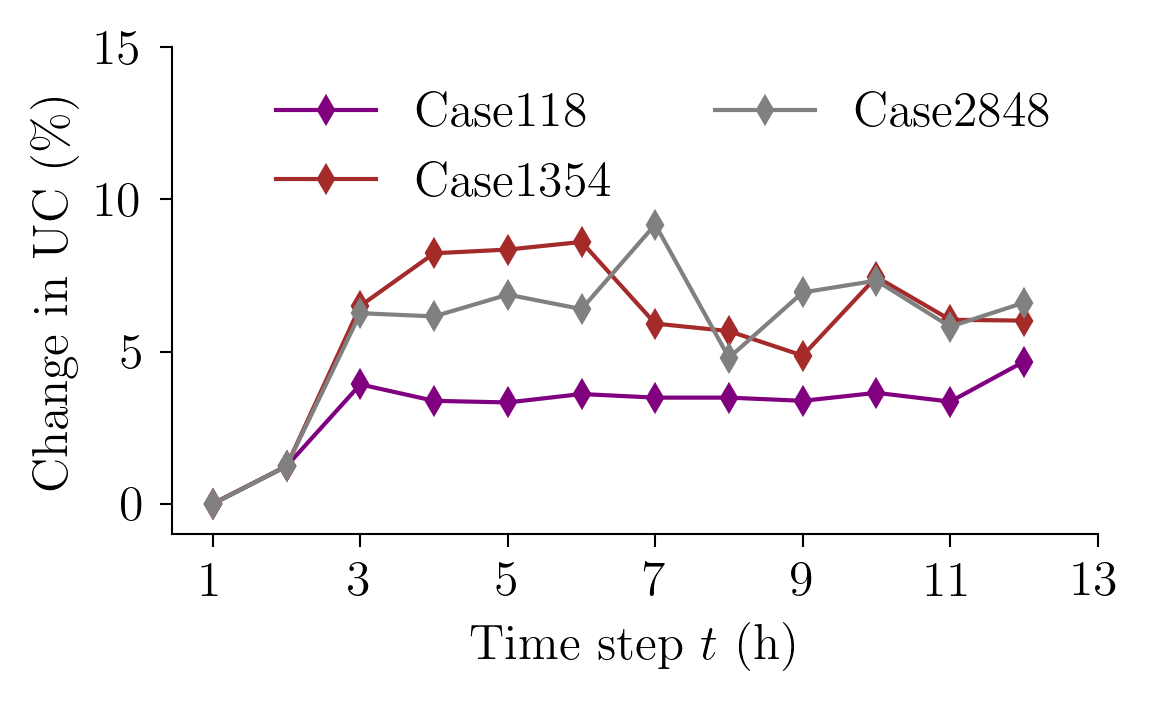}
    \caption{Percentage change in unit commitment (grid topology) over the RUC horizon of interest (i.e., percentage of generators that change status from `ON to OFF' or from `OFF to ON' at each time step)}
    \label{fig:UC_change}
\end{figure}

The average change in the grid topology (i.e., percentage of thermal generators turned on or off) over the RUC horizon, for the testing data points, is shown in Fig.~\ref{fig:UC_change}. All thermal generators are online (status: ON) at $t=1$~h and some of them are turned off at $t=2$~h, because of low anticipated load demand. Thereafter, the forecast scenarios dictate when a generator will be turned on or off. A significant change in thermal generator on-off status proportionately impacts thermal power generation, branch power flow, etc. The results in Fig.~\ref{fig:UC_change} demonstrate that the testing data samples show significant change in the grid topology over the RUC horizon. The risk quantification results under such evolving grid topology are of interest in this work. 

\subsection{GNN model predictions}

In this section, we evaluate the accuracy of the GNN model for system-level, zone-level, and branch-level  QoI predictions.

\begin{figure*}[!ht]
    \centering
    \begin{subfigure}{0.3\linewidth}
        \includegraphics[width=\linewidth]{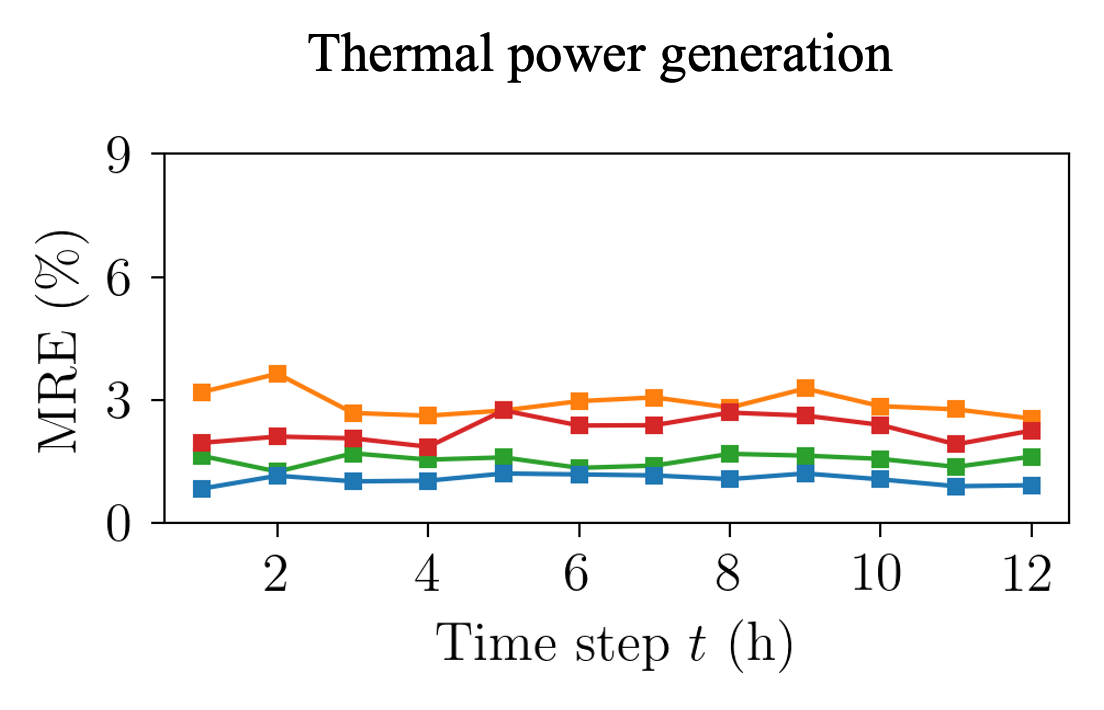}
        \caption{}
        \label{fig:118_agg_PG_MRE}
    \end{subfigure}
    \begin{subfigure}{0.3\linewidth}
        \includegraphics[width=\linewidth]{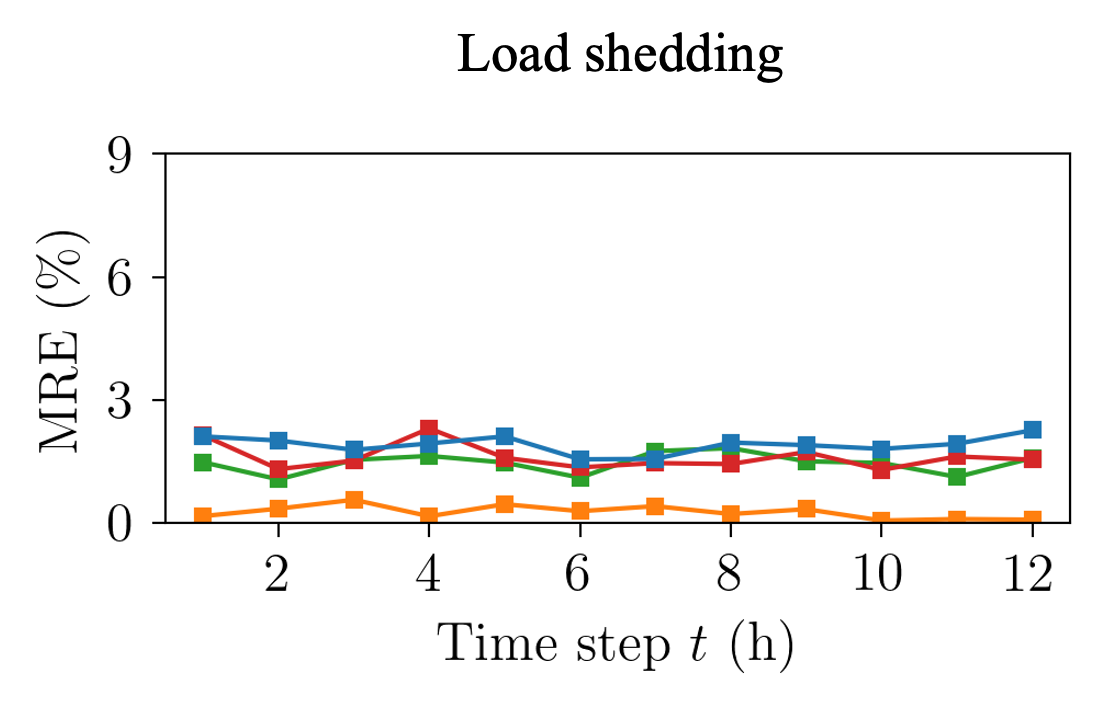}
        \caption{}
        \label{fig:118_agg_Shedding_MRE}
    \end{subfigure}
    \begin{subfigure}{0.3\linewidth}
        \includegraphics[width=\linewidth]{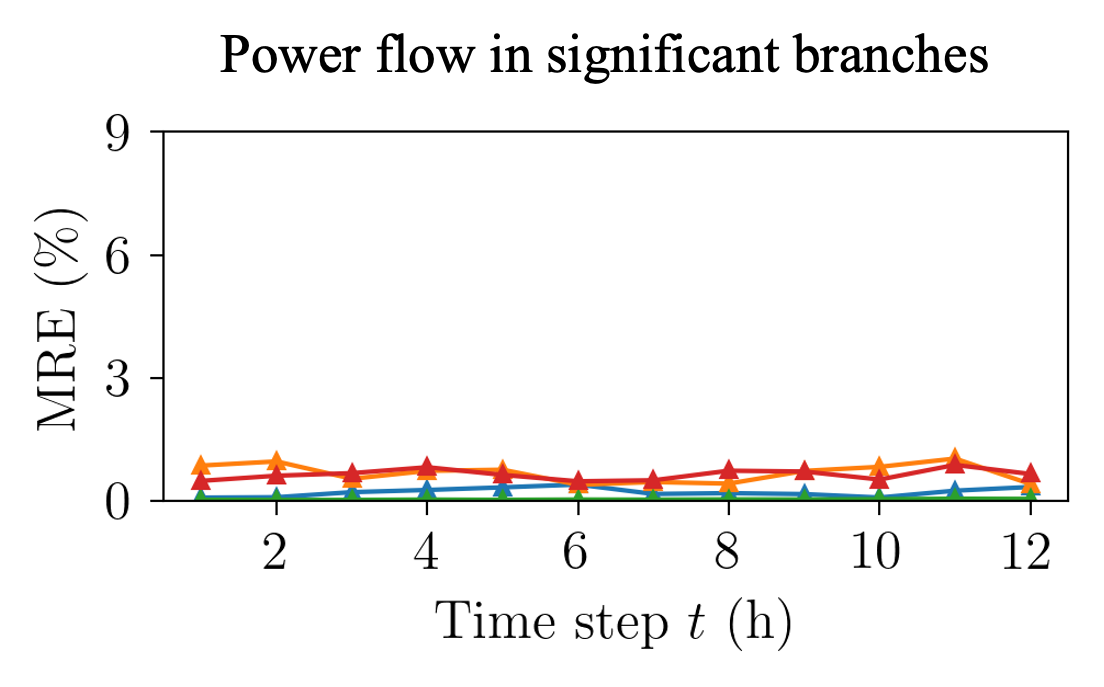}
        \caption{}
        \label{fig:118_PF_MRE}
    \end{subfigure}
    \begin{subfigure}{0.3\linewidth}
        \includegraphics[width=\linewidth]{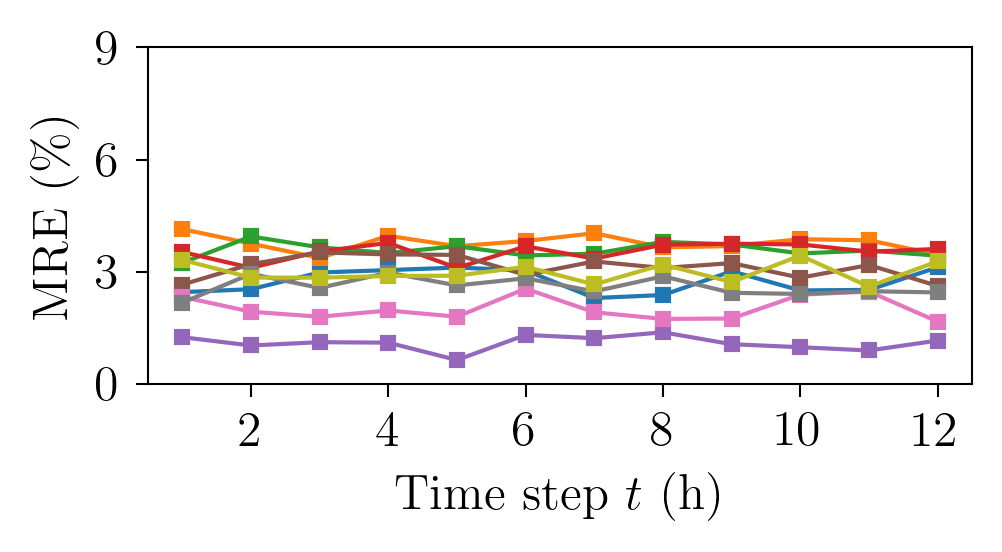}
        \caption{}
        \label{fig:1354_agg_PG_MRE}
    \end{subfigure}
    \begin{subfigure}{0.3\linewidth}
        \centering
        \includegraphics[width=\linewidth]{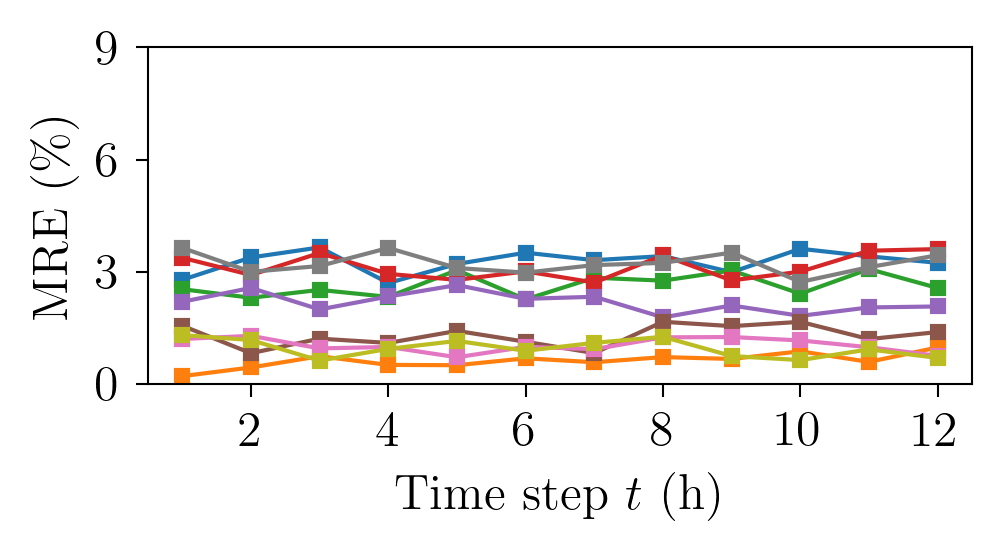}
        \caption{}
        \label{fig:1354_agg_Shedding_MRE}
    \end{subfigure}
    \begin{subfigure}{0.3\linewidth}
        \centering
        \includegraphics[width=\linewidth]{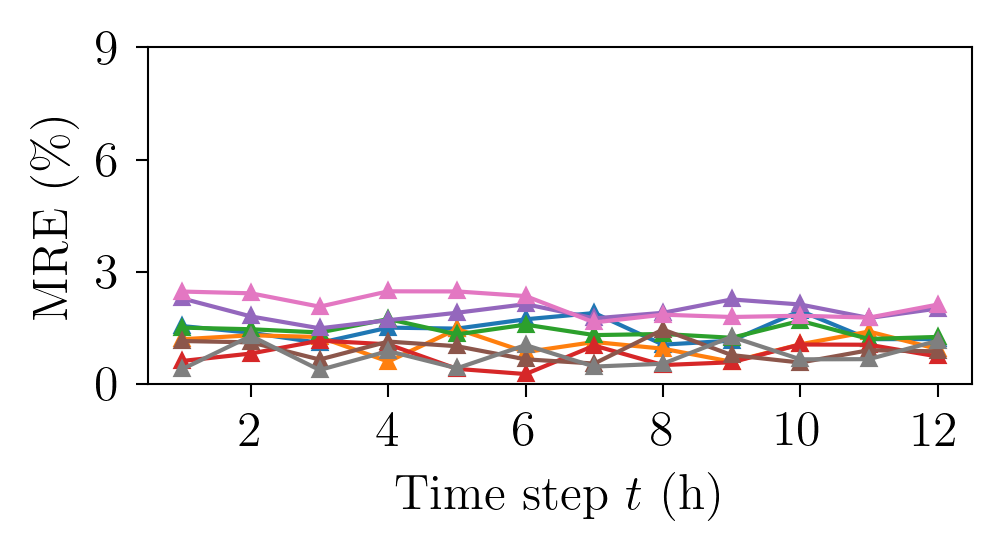}
        \caption{}
        \label{fig:1354_PF_MRE}
    \end{subfigure}
    \begin{subfigure}{0.3\linewidth}
        \includegraphics[width=\linewidth]{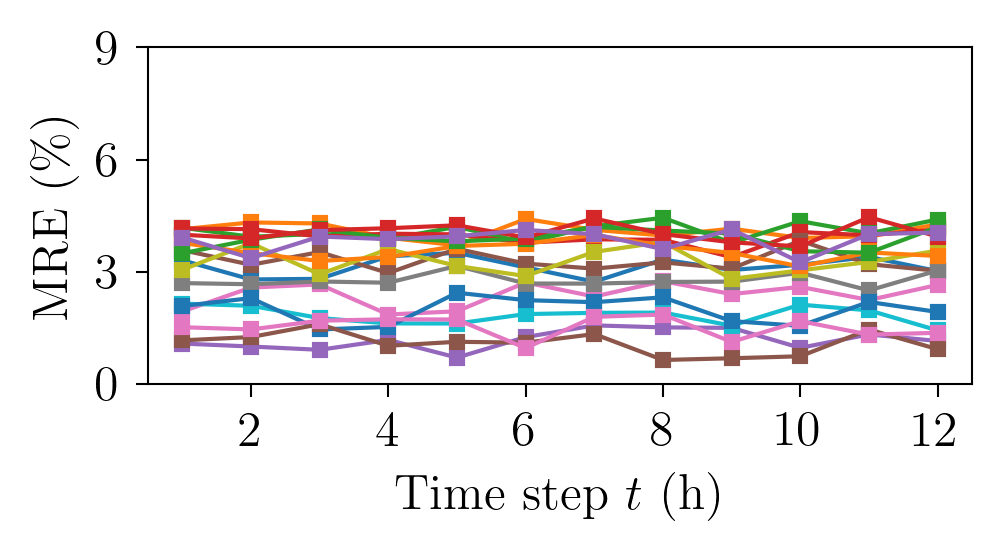}
        \caption{}
        \label{fig:2848_agg_PG_MRE}
    \end{subfigure}
    \begin{subfigure}{0.3\linewidth}
        \centering
        \includegraphics[width=\linewidth]{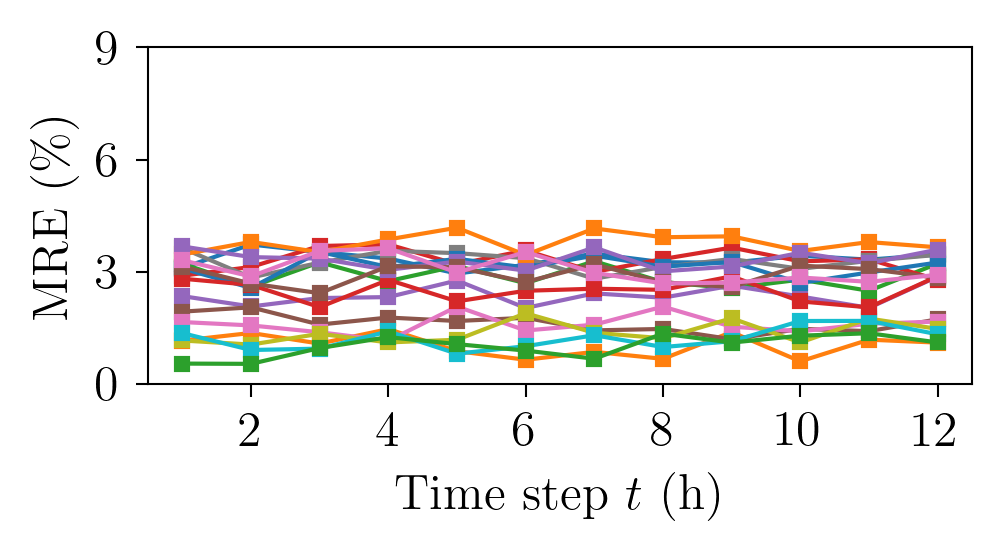}
        \caption{}
        \label{fig:2848_agg_Shedding_MRE}
    \end{subfigure}
    \begin{subfigure}{0.3\linewidth}
        \centering
        \includegraphics[width=\linewidth]{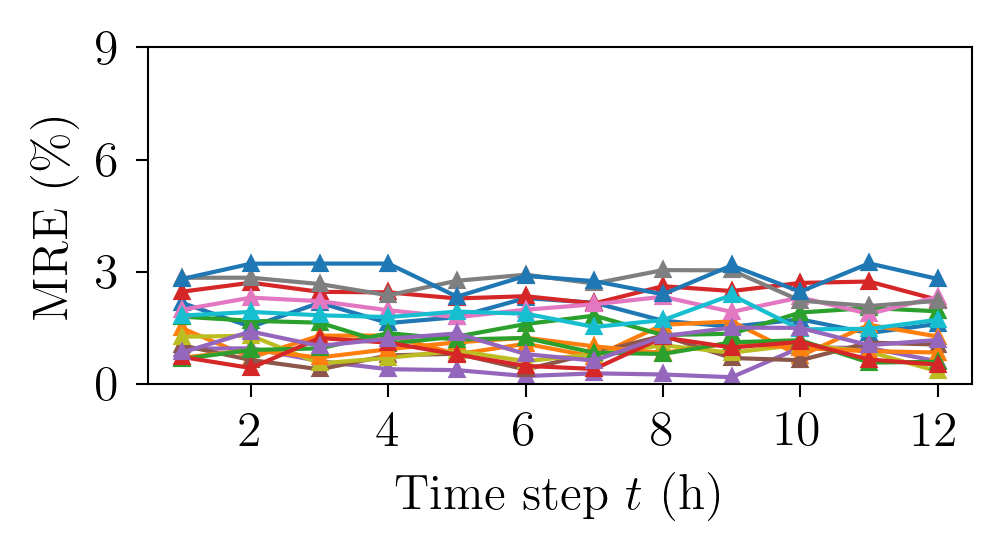}
        \caption{}
        \label{fig:2848_PF_MRE}
    \end{subfigure}
    \begin{subfigure}{0.6\linewidth}
        \centering
        \includegraphics[width=\linewidth]{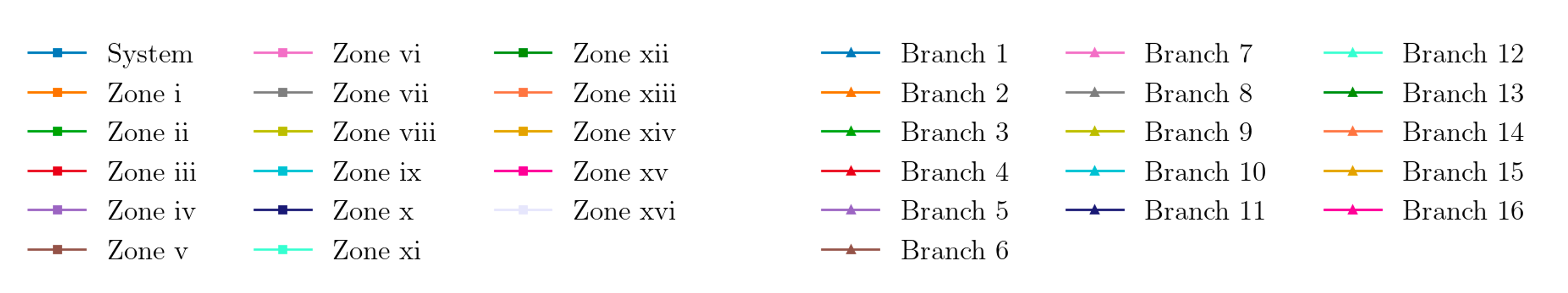}
    \end{subfigure}
    \caption{Mean relative error of GNN predictions for aggregated thermal power generation, aggregated load shedding, and power flow in significant branches. (a)--(c): Case118, (d)--(f): Case1354pegase and (g)--(i): Case2848rte. The number of zones for the three grids are 3, 8 and 16, and the number of (selected) significant branches are 4, 8 and 16, respectively. The same color scheme applies to all subsequent figures unless otherwise specified.}
    \label{fig:MRE}
\end{figure*}

\subsubsection{System-level and zone-level prediction}

For Case118, as shown in Fig.~\ref{fig:118_agg_PG_MRE}, the MRE of aggregated power generation at each time step within the RUC horizon remains below 4\%, indicating excellent accuracy. The MRE for system and zone II is below 2\%. In zone I and III, the MRE is slightly higher but still maintains a good accuracy level with MRE mostly under 3\%; the MRE goes beyond 3\% at two future time steps. The GNN prediction for load shedding is shown in Fig.~\ref{fig:118_agg_Shedding_MRE}. The load shedding prediction is also extremely accurate with the MRE less than 1\% for zone I and 3\% for other zones as well as the system. 

The results for Case1354pegase are summarized in Fig~\ref{fig:1354_agg_PG_MRE}~and~\ref{fig:1354_agg_Shedding_MRE}, results for Case2848rte are shown in Fig.~\ref{fig:2848_agg_PG_MRE}~and~\ref{fig:2848_agg_Shedding_MRE}, respectively. Even for these large grids GNN predictions exhibit good accuracy  (MRE $<$5\%) for all zones and the system at different time steps.

\subsubsection{Branch-level prediction}

In terms of the GNN prediction for branch flow, the results are shown in Fig.~\ref{fig:118_PF_MRE}. Note that only the results for significant branches are shown here. It is noticed that MRE is consistently below 1\% for Case118, suggesting excellent predictive accuracy of the GNN model. The results for the other two grids are shown in Fig~\ref{fig:1354_PF_MRE}~and~\ref{fig:2848_PF_MRE}, and GNN prediction MRE remains below 3\% for these large grids. For a comprehensive evaluation, the maximum relative error of GNN prediction is also provided in~\ref{sec:max_prediction_error}.

Overall, the GNN surrogates achieve excellent accuracy in predicting zonal/system level (aggregated) and branch-level quantities. On average, a numerical MILP solver requires $12 \, s$, $78 \, s$ and $175 \, s$, respectively, to solve one SCUC problem for the three grids. It only takes the GNN model $2 \times 10^{-4} \, s$, $10^{-3} \, s$ and $1.5 \times 10^{-3} \, s$ respectively to obtain the corresponding prediction. The speed-up achieved using the GNN surrogate model is thus capable of enabling real-time risk quantification with adequate accuracy.

\subsection{Reliability and risk predictions} 

In this section, we evaluate the accuracy of GNN-based reliability and risk assessment by comparing it with MILP-based assessment.

\subsubsection{Load shedding}

\begin{figure*}[!ht]
    \centering
    \begin{subfigure}{0.24\linewidth}
        \centering
        \includegraphics[width=\linewidth]{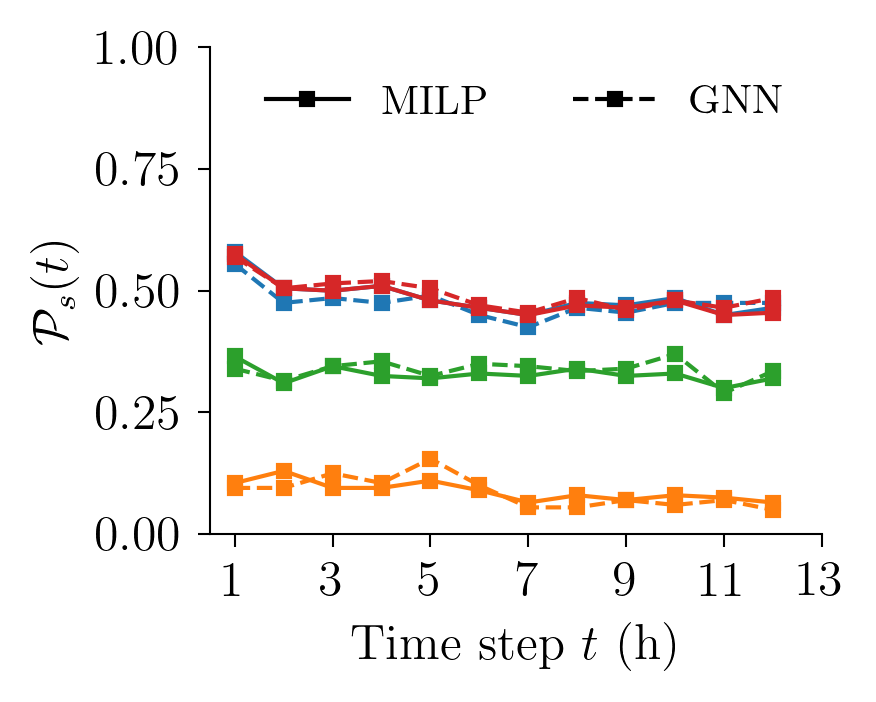}
        \caption{}
        \label{fig:118_shedding_standalone_reliability}
    \end{subfigure}
    \hfill
    \begin{subfigure}{0.24\linewidth}
        \centering
        \includegraphics[width=\linewidth]{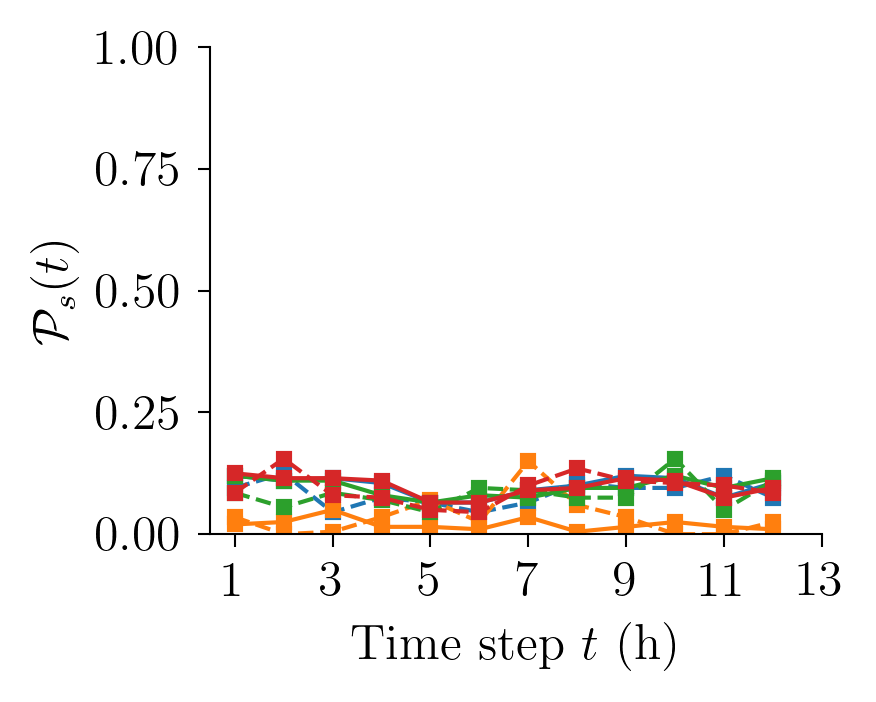}
        \caption{}
        \label{fig:118_shedding_standalone_reliability_reserve}
    \end{subfigure}
    \hfill
    \begin{subfigure}{0.24\linewidth}
        \centering
        \includegraphics[width=\linewidth]{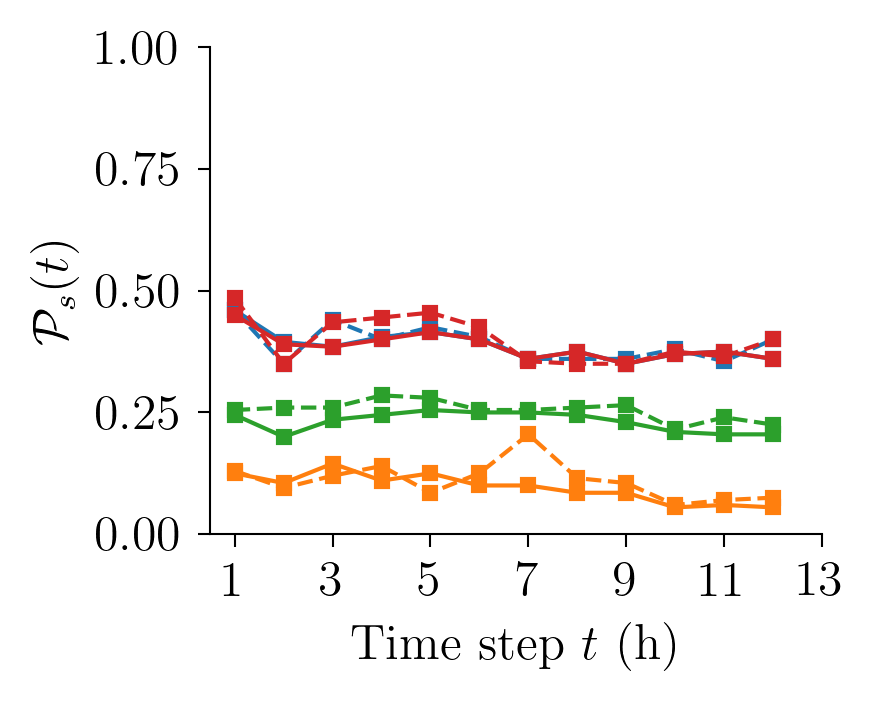}
        \caption{}
        \label{fig:118_shedding_standalone_reliability_no_reserve}
    \end{subfigure}
    \begin{subfigure}{0.24\linewidth}
        \centering
        \includegraphics[width=\linewidth]{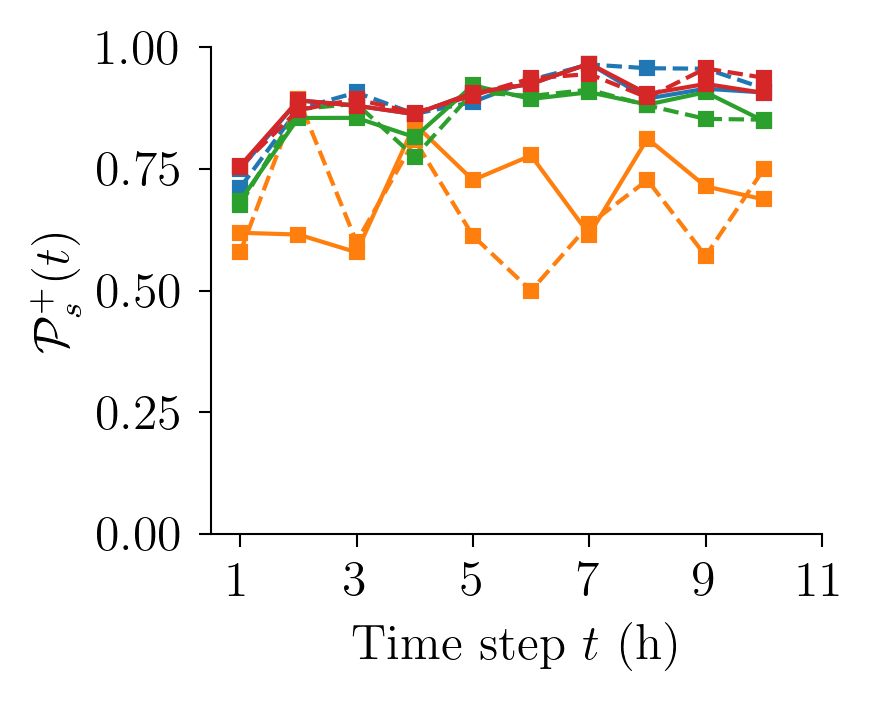}
        \caption{}
        \label{fig:118_shedding_multi-step_reliability}
    \end{subfigure}
    \label{fig:118_shedding_reliability}
    \caption{Reliability and risk quantification for load shedding (Case118). (a) Probability of (standalone) total (reserve- and non-reserve-related) load shedding, (b) probability of (standalone) load shedding due to reserve constraint, (c) probability of (standalone) load shedding due to non-reserve constraints, and (d) probability of (multi-step) total (reserve- and non-reserve-related) load shedding.}
\end{figure*}

\begin{figure}
    \centering
    \includegraphics[width=0.5\linewidth]{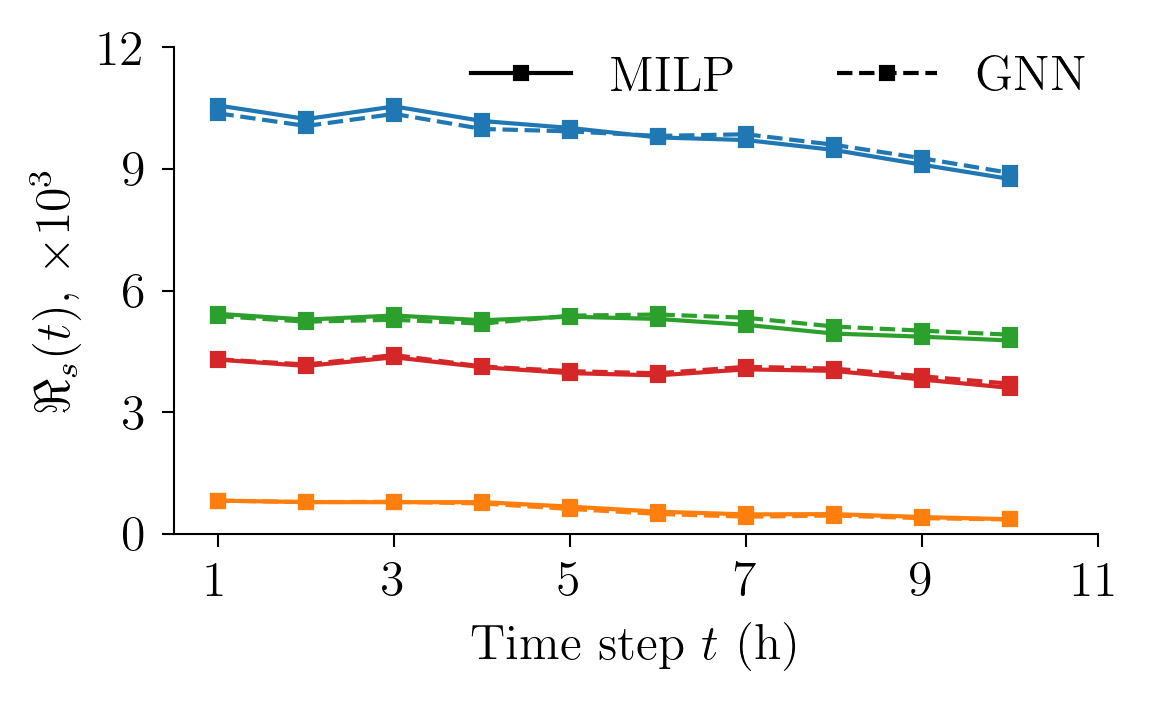}
    \caption{Risk of load shedding (Case118).}
    \label{fig:118_shedding_overall_risk}
\end{figure}

The probability of load shedding for Case118 is shown in Fig.~\ref{fig:118_shedding_standalone_reliability}. GNN-based reliability quantification is in agreement with the ground truth (MILP solver-based quantification), for both the system and individual zones. The probabilities of standalone total (reserve- plus non-reserve-related) load shedding at zones I, II and III are around 0.1, 0.4 and 0.5, respectively, suggesting that zone II and III have higher vulnerability than zone I. At the system level, the probability of load shedding is around 0.5 over the planning horizon. Load shedding probabilities w.r.t. the reserve requirement and other constraints are also estimated and shown in  Fig.~\ref{fig:118_shedding_standalone_reliability}~and~\ref{fig:118_shedding_standalone_reliability_no_reserve}. The GNN model shows excellent predictive capability for cause-aware load shedding analysis. Unmet reserve requirement constraint seems to be the primary cause of load shedding, as the corresponding probability is significantly larger than that corresponding to non-reserve constraints. The GNN-based predictions could be used to analyze and mitigate the reserve requirement-related load shedding during the subsequent few hours. The results suggest that GNN is capable of decoupling different factors that lead to the adverse event, thus it can be a useful tool to improve the operator’s situational awareness and enhance the system reliability. The multi-step reliability quantification is shown in Fig.~\ref{fig:118_shedding_multi-step_reliability}. It is seen that GNN-based reliability calculation is very accurate (except for Zone I) and captures the trend of load shedding probability in future two-hour-long windows starting at each of the future hours of interest.

Not surprisingly, GNN-based risk quantification aligns well with that from the reference (MILP-based) solution, as shown in Fig.~\ref{fig:118_shedding_overall_risk}. It is observed that the risk caused by load shedding gradually decreases. This is reasonable as risk is estimated at the current time step, the farther it goes into future, the more time left for operators to take action. We also notice that zone I is significantly less risky than other zones, and the consequence cost almost drops to zero after $T = 8$ hours. In contrast, zone II exhibits the highest risk at around $\$$5,000, slightly higher than that of zone III which is at $\$$4,000. The system-level risk caused by load shedding experiences a slight drop from around $\$$11,000 at $t=1$~h to $\$$9,000 at $t=12$~h, and the trend is accurately captured by the GNN-based risk computation. 

\begin{figure*}[!ht]
    \centering
    \begin{subfigure}{0.32\linewidth}
        \centering
        \includegraphics[width=\linewidth]{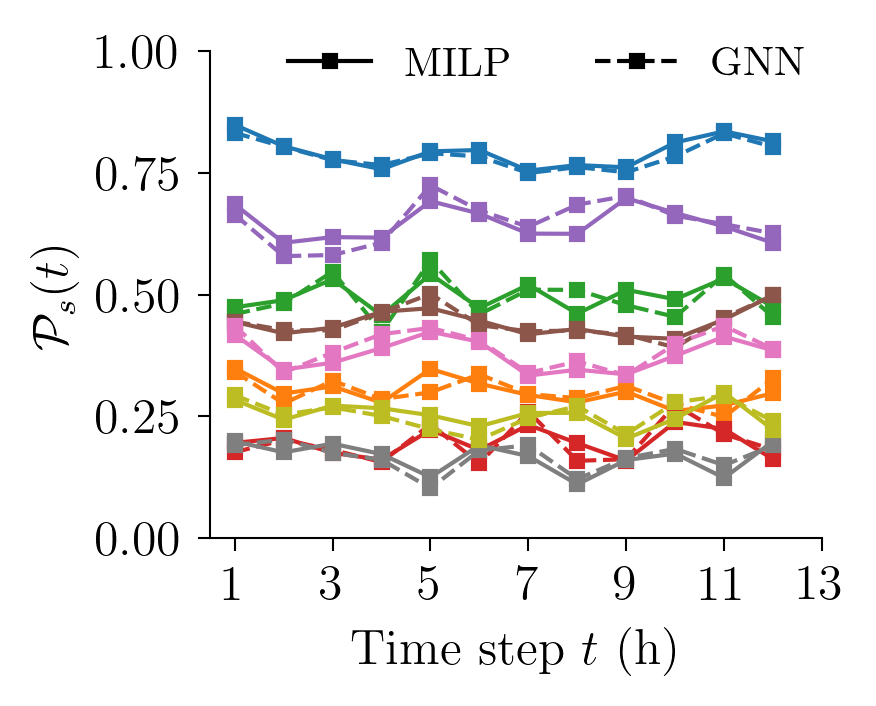}
        \caption{}
        \label{fig:1354_shedding_standalone_reliability}
    \end{subfigure}
    \hfill
    \begin{subfigure}{0.32\linewidth}
        \centering
        \includegraphics[width=\linewidth]{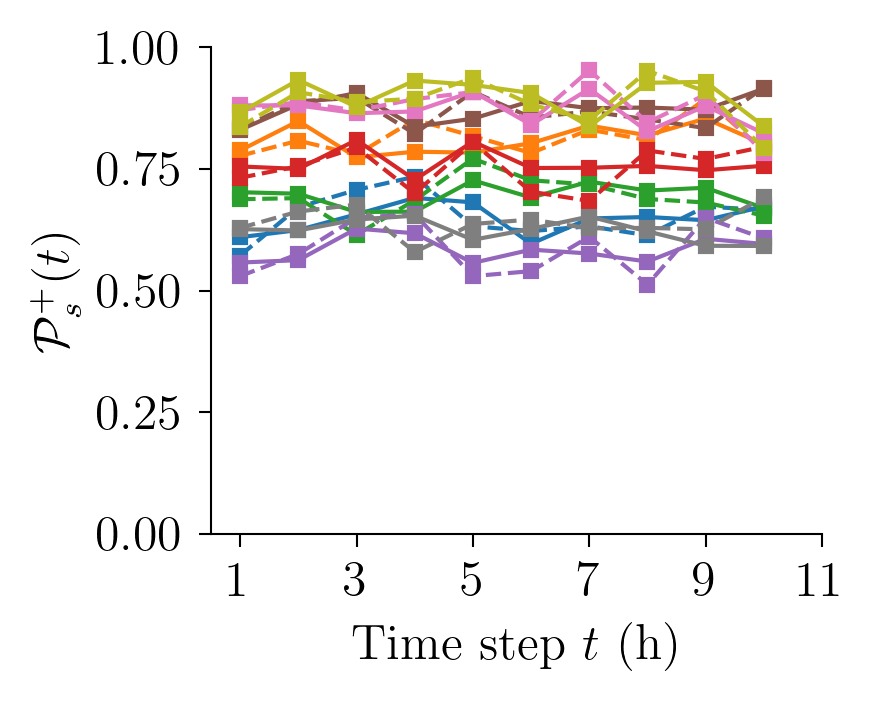}
        \caption{}
        \label{fig:1354_shedding_multi-step_reliability}
    \end{subfigure}
    \hfill
    \begin{subfigure}{0.32\linewidth}
        \centering
        \includegraphics[width=\linewidth]{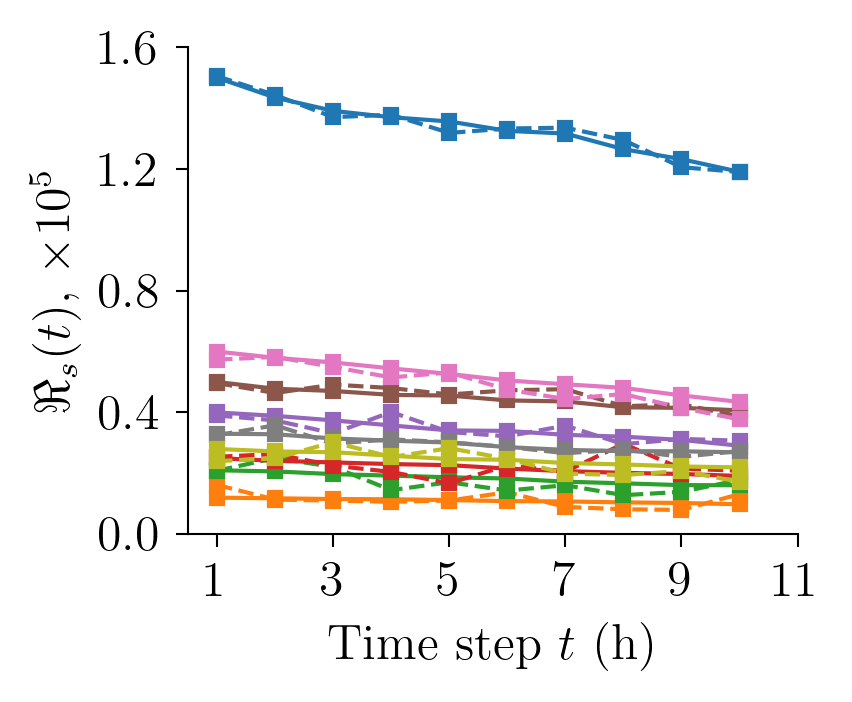}
        \caption{}
        \label{fig:1354_shedding_overall_risk}
    \end{subfigure}
    \caption{Reliability and risk quantification for load shedding (Case1354pegase). (a) Probability of standalone load shedding, (b) probability of multi-step load shedding, and (c) risk of load shedding.}
    \label{fig:1354_shedding_RR}
\end{figure*}

\begin{figure*}[!ht]
    \centering
    \begin{subfigure}{0.32\linewidth}
        \centering
        \includegraphics[width=\linewidth]{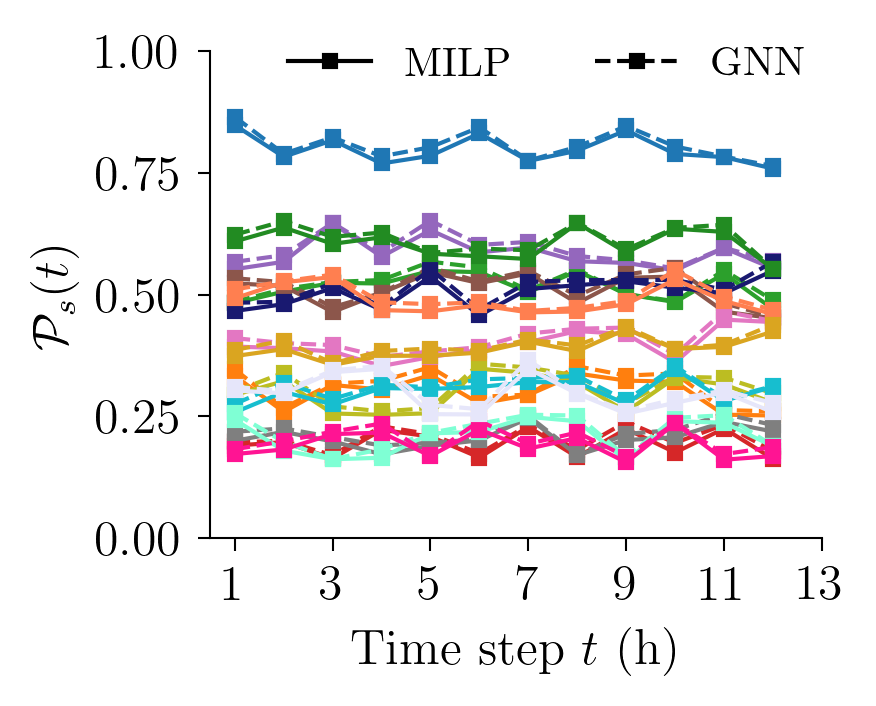}
        \caption{}
        \label{fig:2848_shedding_standalone_reliability}
    \end{subfigure}
    \hfill
    \begin{subfigure}{0.32\linewidth}
        \centering
        \includegraphics[width=\linewidth]{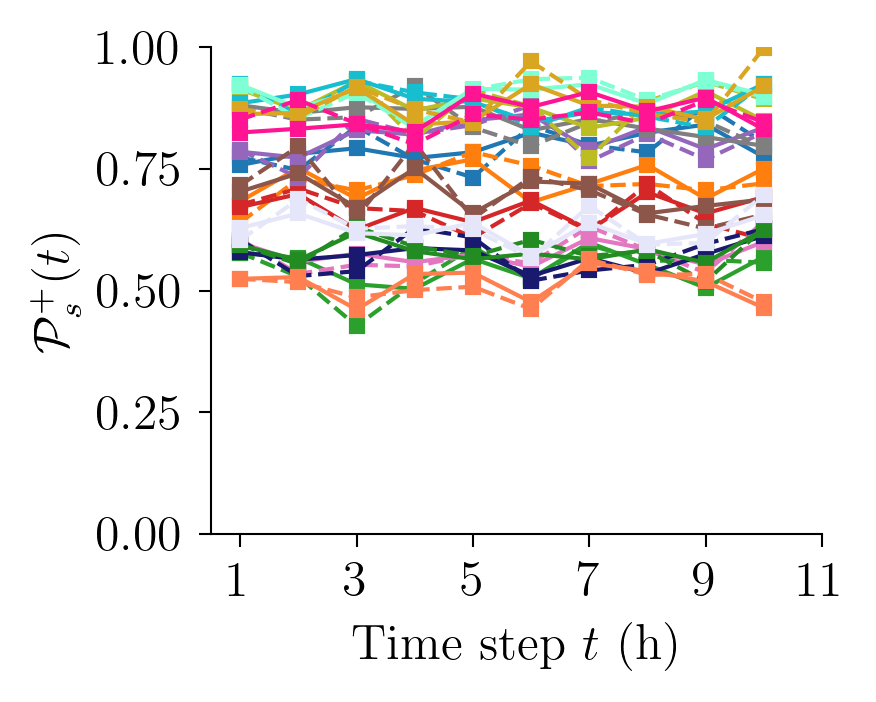}
        \caption{}
        \label{fig:2848_shedding_multi-step_reliability}
    \end{subfigure}
    \hfill
    \begin{subfigure}{0.32\linewidth}
        \centering
        \includegraphics[width=\linewidth]{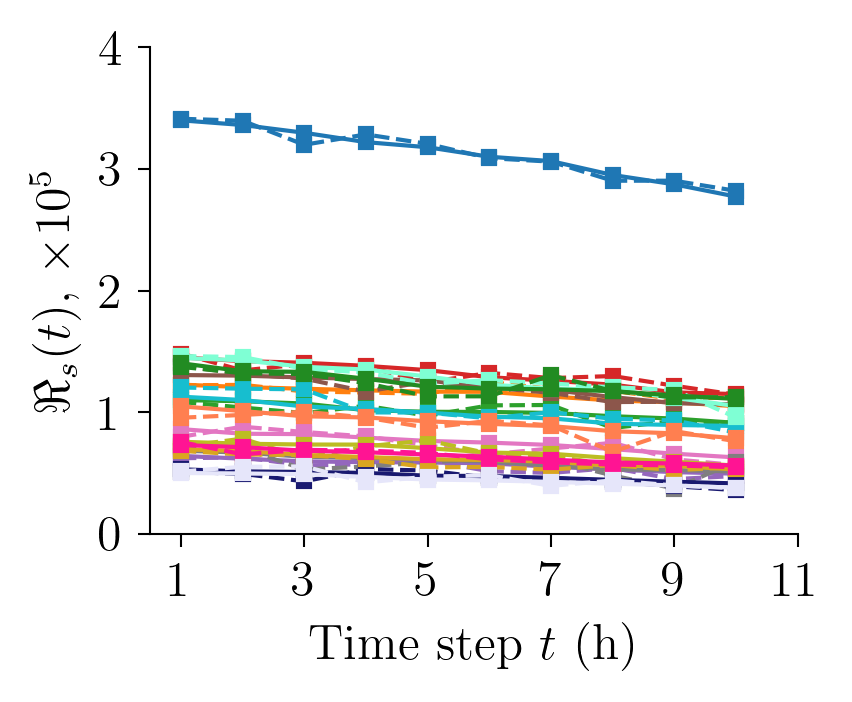}
        \caption{}
        \label{fig:2848_shedding_overall_risk}
    \end{subfigure}
    \caption{Reliability and risk quantification for load shedding (Case2848rte). (a) Probability of standalone load shedding, (b) probability of multi-step load shedding, and (c) risk of load shedding.}
    \label{fig:2848_shedding_RR}
\end{figure*}

The reliability and risk quantification of load shedding for Case1354pegase and Case2848rte are shown in Fig.~\ref{fig:1354_shedding_RR}~and~\ref{fig:2848_shedding_RR}, respectively. It can be seen that GNN-based risk quantification is in excellent agreement with the reference (MILP-based) solution for individual zones as well as for the entire system. The results suggest that GNN can be used to facilitate risk management of large, real-world grids by speeding up the computation. Since well-trained GNN models are able to give QoI predictions for thousands of operational scenarios within a second, the GNN-based approach is thus appealing for hours-ahead reliability and risk assessment and management.

\subsubsection{Branch overloading}

\begin{figure*}[!ht]
    \centering
    \begin{subfigure}{0.32\linewidth}
        \centering
        \includegraphics[width=\linewidth]{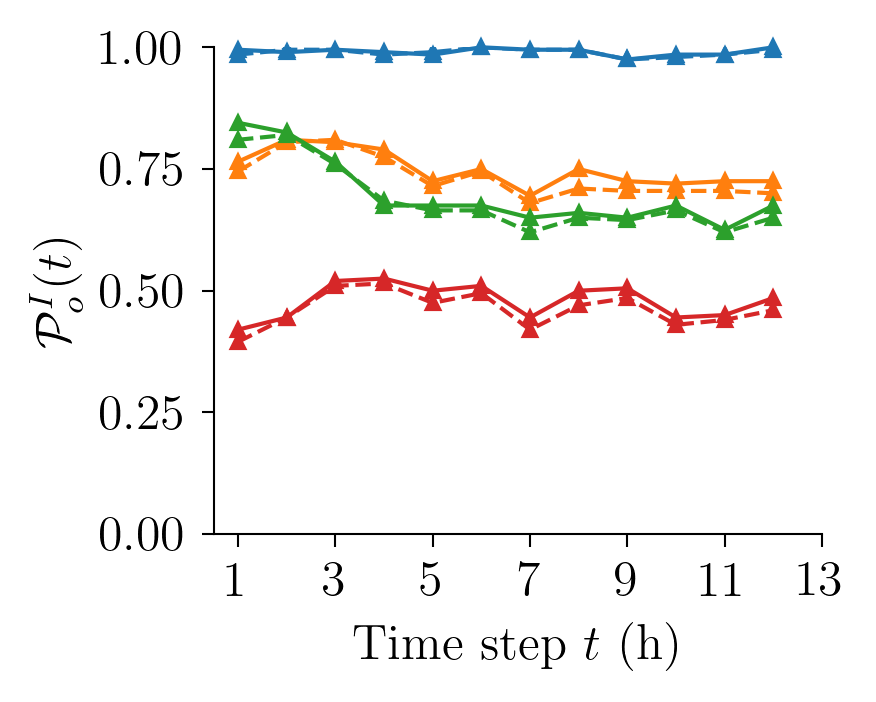}
        \caption{}
        \label{fig:118_overloading_standalone_reliability}
    \end{subfigure}
    \begin{subfigure}{0.32\linewidth}
        \centering
        \includegraphics[width=\linewidth]{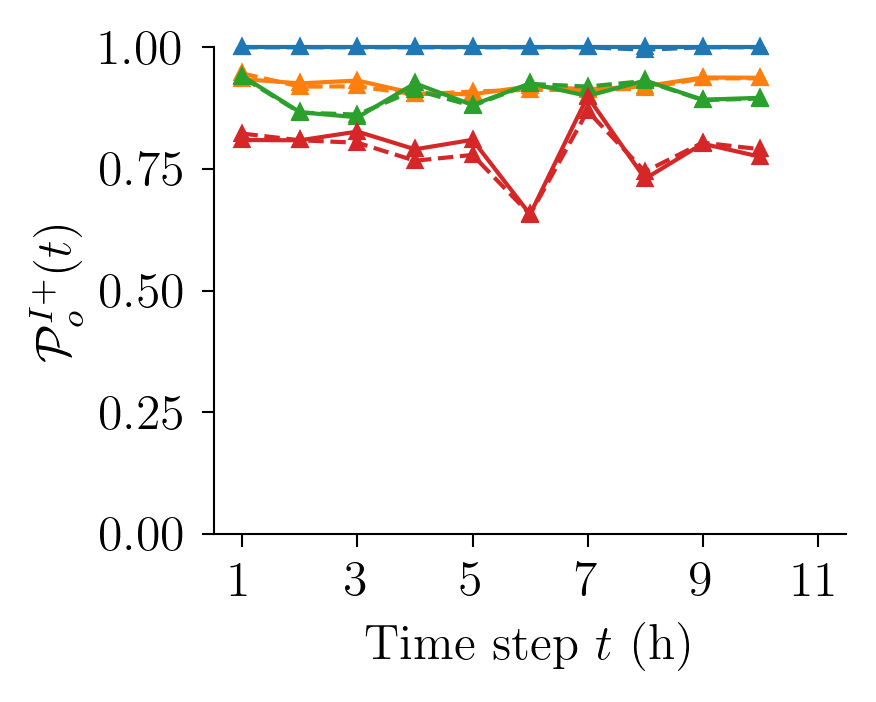}
        \caption{}
        \label{fig:118_overloading_multi-step_reliability}
    \end{subfigure}
    \begin{subfigure}{0.32\linewidth}
        \centering
        \includegraphics[width=\linewidth]{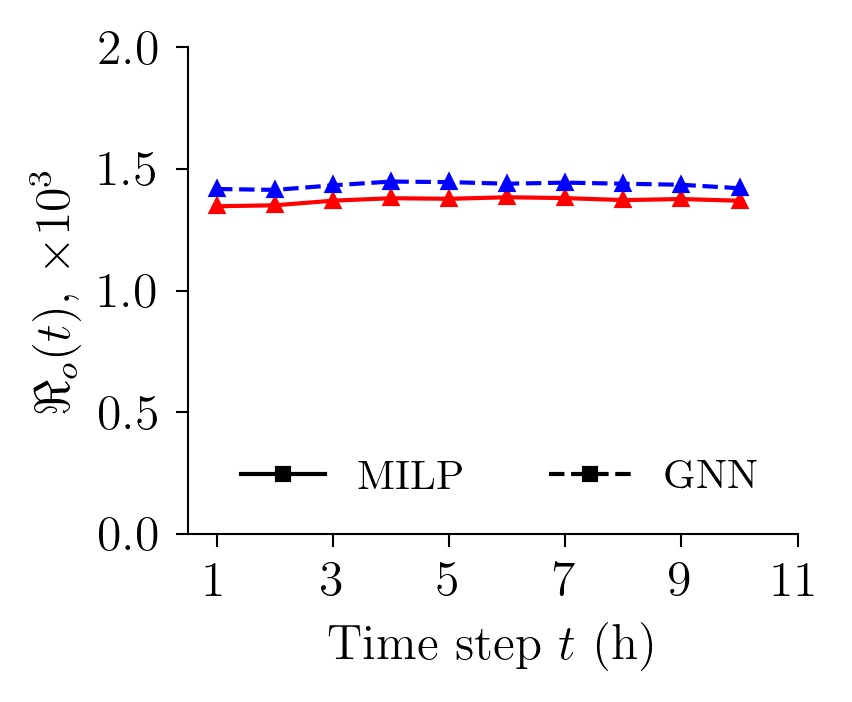}
        \caption{}
        \label{fig:118_overloading_overall_risk}
    \end{subfigure}
    \caption{Reliability and risk of branch overloading (Case118). (a) Probability of standalone branch overloading, (b) Probability of multi-step branch overloading, and (c) Risk of branch overloading.}
    \label{fig:118_overloading_RR}
\end{figure*} 

The probability of standalone branch overloading is shown in Fig.~\ref{fig:118_overloading_standalone_reliability}. It is observed that the GNN-based analysis for the selected (significant) branches is very accurate. We notice that the first branch is almost always overloaded with probability near 1 at all time steps, and the second and third branches are often overloaded with probabilities around 0.7 and 0.6, respectively. The fourth branch is overloaded with a  probability of 0.5. The results also demonstrate that it is not necessary to include all the branches in reliability and risk assessment, since the overloading probability for the remaining the branches is  smaller than 0.5. 

The probabilities of multi-step branch overloading are shown in Fig.~\ref{fig:118_overloading_multi-step_reliability}. Compared with standalone overloading, the probability of multi-step overloading for all branches is slightly elevated. In particular, the first branch is still the most vulnerable one with the probability of overloading at 1 at all time steps. For the second and third branch, multi-step overloading probability is elevated to 0.9 and 0.8, respectively. Overloading probability for the fourth branch increases to around 0.7. The risk associated with branch overloading is shown in Fig.~\ref{fig:118_overloading_overall_risk}. It is observed that the GNN-based risk quantification is slightly conservative compared to the reference (MILP-based) solution (however, the error is negligibly small). 

\begin{figure*}
    \centering
    \begin{subfigure}{0.32\linewidth}
        \centering
        \includegraphics[width=\linewidth]{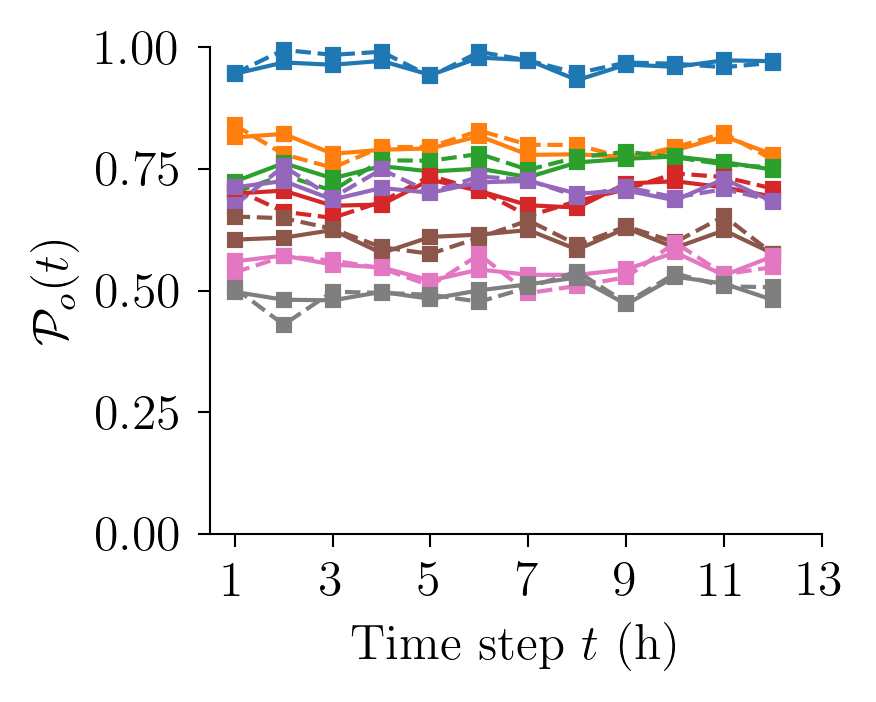}
        \caption{}
        \label{fig:1354_overloading_standalone_reliability}
    \end{subfigure}
    \begin{subfigure}{0.32\linewidth}
        \centering
        \includegraphics[width=\linewidth]{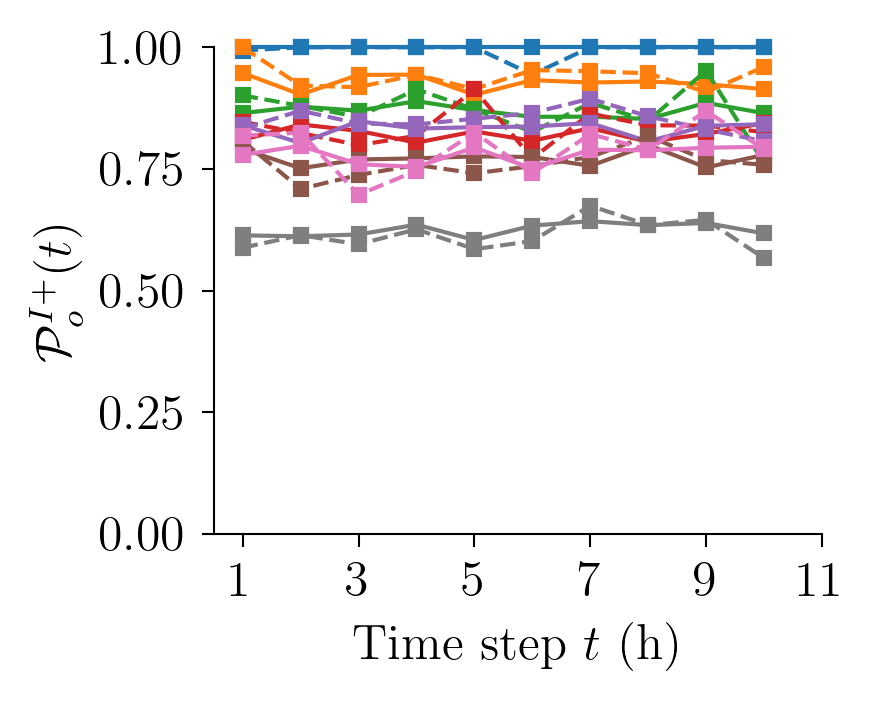}
        \caption{}
        \label{fig:1354_overloading_multi-step_reliability}
    \end{subfigure}
    \begin{subfigure}{0.32\linewidth}
        \centering
        \includegraphics[width=\linewidth]{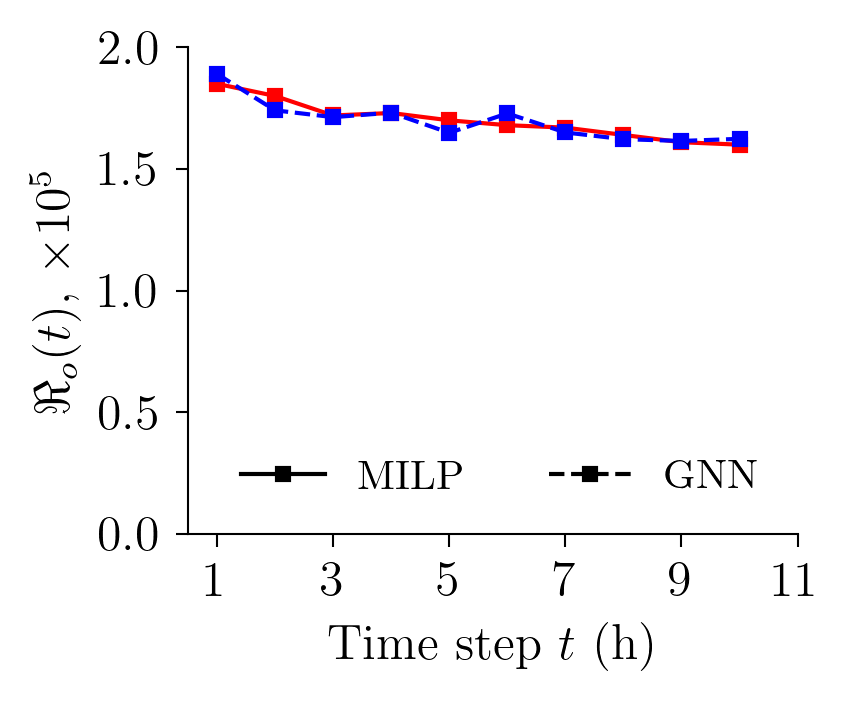}
        \caption{}
        \label{fig:1354_overloading_overall_risk}
    \end{subfigure}
    \caption{Reliability and risk of branch overloading (Case1354pegase). (a) Probability of standalone branch overloading, (b) Probability of multi-step branch overloading, and (c) Risk of branch overloading.}
    \label{fig:1354_overloading_RR}
\end{figure*}

\begin{figure*}
    \centering
    \begin{subfigure}{0.32\linewidth}
        \centering
        \includegraphics[width=\linewidth]{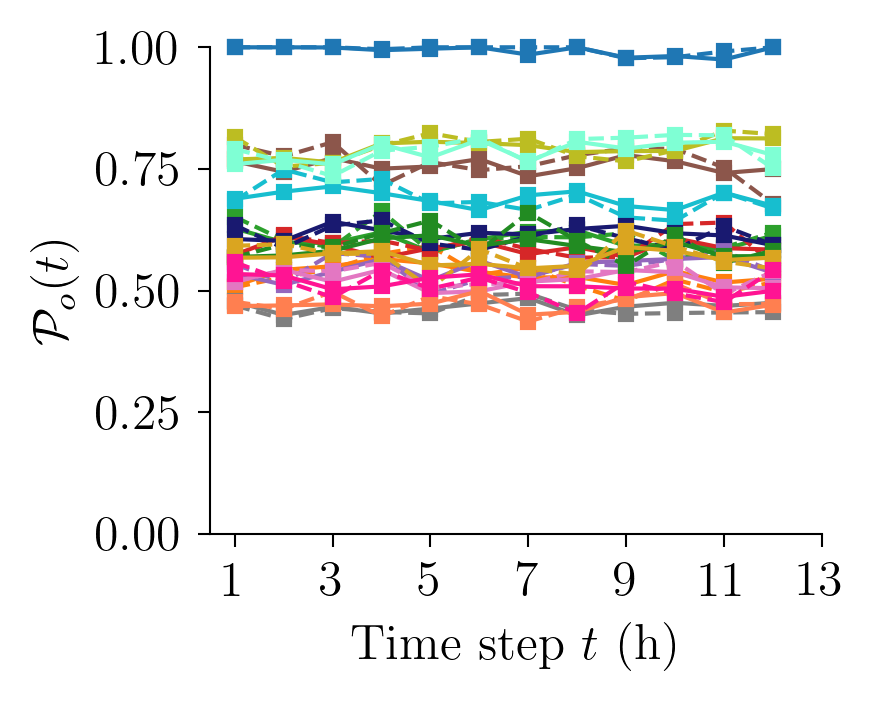}
        \caption{}
        \label{fig:2848_overloading_standalone_reliability}
    \end{subfigure}
    \begin{subfigure}{0.32\linewidth}
        \centering
        \includegraphics[width=\linewidth]{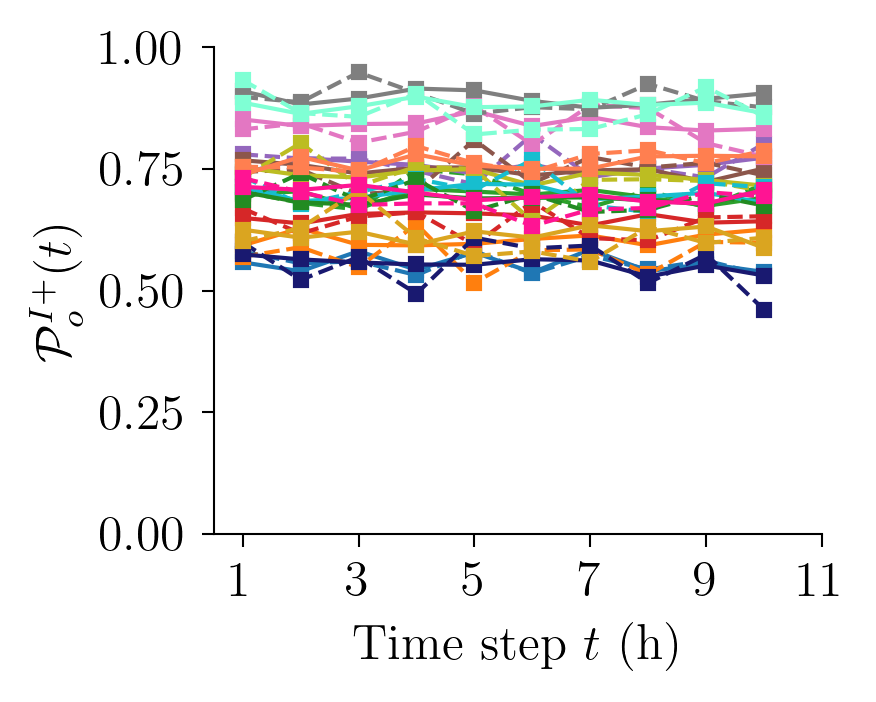}
        \caption{}
        \label{fig:2848_overloading_multi-step_reliability}
    \end{subfigure}
    \begin{subfigure}{0.32\linewidth}
        \centering
        \includegraphics[width=\linewidth]{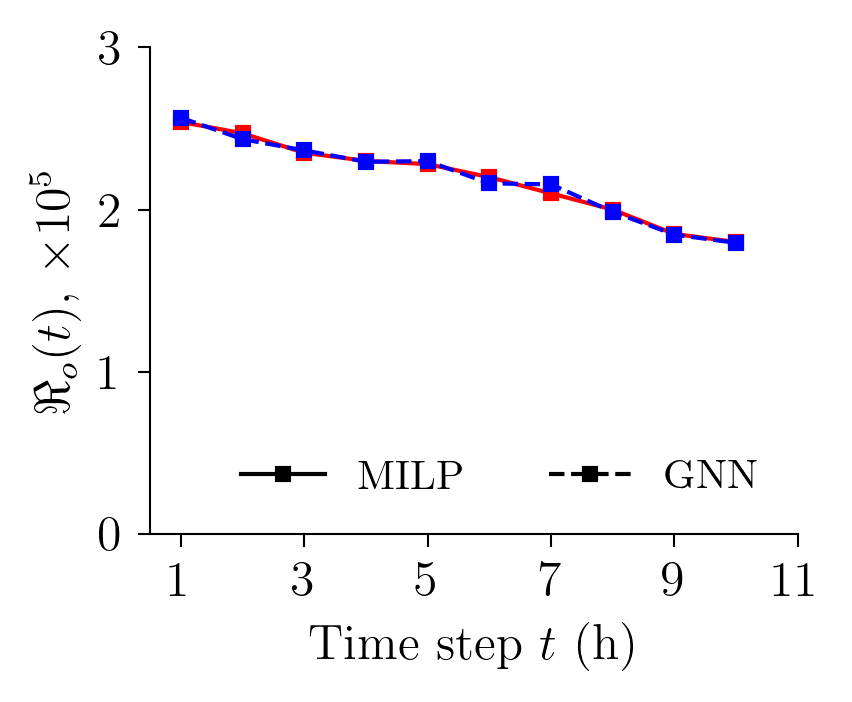}
        \caption{}
        \label{fig:2848_overloading_overall_risk}
    \end{subfigure}
    \caption{Reliability and risk of branch overloading (Case2848rte). (a) Probability of standalone branch overloading, (b) Probability of multi-step branch overloading, and (c) Risk of branch overloading.}
    \label{fig:2848_overloading_RR}
\end{figure*}

The reliability and risk quantification results of branch overloading for Case1354pegase and Case2848rte are shown in Fig.~\ref{fig:1354_overloading_RR}~and~\ref{fig:2848_overloading_RR}, respectively, and the results confirm the good accuracy of GNN-based results compared to MILP-based results. This demonstrates the excellent accuracy of GNN-based risk quantification; hence the GNN surrogate model can be a valuable tool to enable fast risk quantification during  power grid operation, thus supporting fast, risk-informed decision-making.

% #########################################################################
% ############################### Conclusion ##############################
% #########################################################################
\section{Conclusion}
\label{sec:conclusion}

We investigated the utilization of graph neural network (GNN) as a proxy to facilitate hours-ahead decision-making in power grid operations, while accounting for anticipated changes in the grid topology (i.e., changes in the generator on/off status). To train the GNN model,  spatio-temporally correlated samples of stochastic variables (wind power and load demand) are drawn from their joint probability (forecast) distribution, and the corresponding unit commitment and dispatch solutions are obtained using an MILP solver. The data (samples and MILP solutions) are used to train the GNN models following a supervised learning approach with stochastic grid variables (wind/solar generation and load)  as inputs and MILP solutions as outputs. Multiple GNN models are trained to predict different quantities of interest (QoIs), i.e., load shedding at system/zonal level and power flow at branch level. The QoIs are then used for reliability and risk quantification using Monte Carlo sampling. Failure modes at system level (load shedding) and branch level (transmission line overloading) are considered. The methodology employs standalone and multi-step temporal views of failure events for hours-ahead risk assessment, and thus provides a comprehensive evaluation of risk associated with adverse events. The proposed GNN-based reliability and risk estimation methodology was demonstrated on medium to large realistic power grids. The excellent prediction accuracy and low computational cost of the GNN model indicate that GNN models can be good proxies for complicated computational tasks and fast decision-making in power grid operation. Our results also demonstrated that GNN models can provide accurate estimate of hours-ahead grid operational risk. 

The proposed GNN-based reliability and risk assessment method utilizes probabilistic forecasts over the next few hours to obtain the training data. The subsequently trained GNN proxies are only applicable to the probabilistic forecasts used for generating the training data. Future work needs to develop GNN proxies that can be generalized to a wide range of possible forecasts. The statistical distance between training data and out-of-distribution testing data could be used to gauge the performance of the GNN proxies by comparing QoI prediction accuracy.

\section*{Acknowledgement}

This work was partly funded by the ARPA-E PERFORM program (Grant no.: DE-AR0001280, Project technical monitor: Jonathan~Glass) under subcontract to Georgia Tech (PI: Prof.~Pascal~Van~Hentenryck). The support is gratefully acknowledged. The authors also acknowledge helpful discussions with Dr.~Oliver~Stover at Charles River Associates and Dr.~Ray~Daniel~Zimmerman at Cornell University.

\doublespacing
\bibliography{refs}

\newpage

\appendix

\section{Synthetic power grids}
\label{sec:power_grids}

% Add "S" in section number
% \renewcommand{\thesection}{S\arabic{section}}
% % Add "S"
% \renewcommand{\thefigure}{S-\arabic{figure}}   
\setcounter{figure}{0}   
\setcounter{table}{0}

\subsection{Case118}

\begin{figure}[H]
    \centering
    \includegraphics[width=0.5\linewidth]{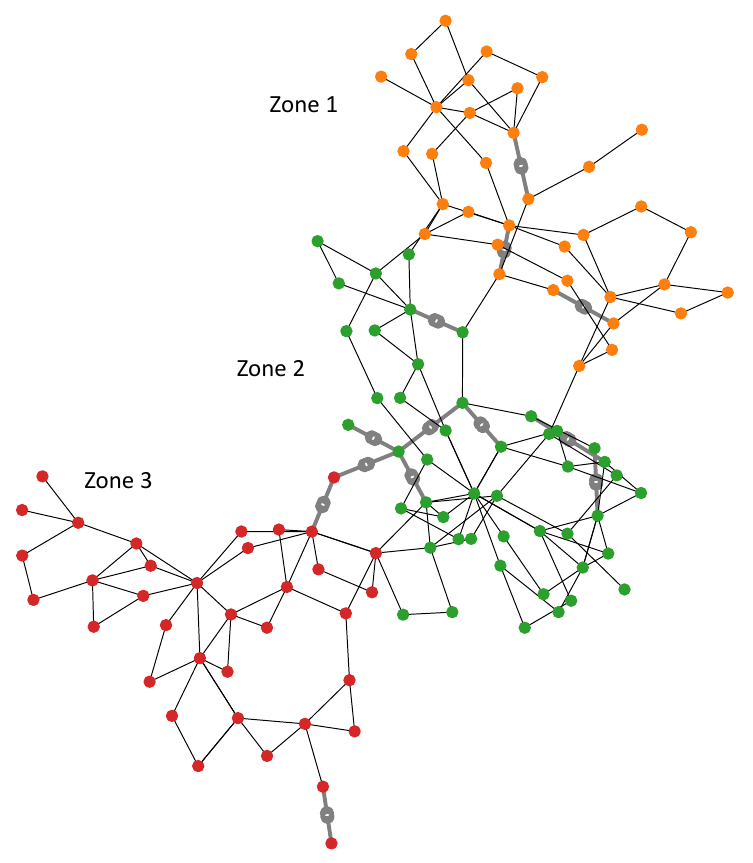}
    \caption{Case118 power grid.}
    \label{fig:Case118_power_grid}
\end{figure}

Case118 power grid contains 118 buses (54 generators, 99 loads and 186 branches). It is partitioned into three zones as shown in Fig.~\ref{fig:Case118_power_grid} and Table~\ref{tab:Case118_zone_info}. For each zone, an aggregated value of load and wind power generation are drawn from their respective probability distributions. The relative contribution of each load bus/wind turbine is assumed to be a constant in each zone. Consider zone $i$ containing $N_L$ load buses and $N_{W}$ wind turbines; then the bus-level load and wind power for zone $i$ are calculated as:
\begin{align} \label{eqn:relative_contribution}
\begin{split}
L_{i, j} &= r_{i, j}L_i, \; j = 1, 2, ..., N_L \\
W_{i, k} &= q_{i, k}W_i, \; k = 1, 2, ..., N_W    
\end{split}
\end{align}
where $r_{i, j}$ and $q_{i, k}$ represent the relative contribution of each load bus and wind generator bus to the respective zonal (aggregated) values, and $L_i$ and $W_i$ denote the aggregated values for zone $i$. It is assumed that load follows truncated normal PDF, whereas wind power generation is converted from wind speed, which is assumed to follow Weibull PDF. The probability distribution parameters are listed in Table~\ref{tab:Case118_PDF_parameters}.

Wind power generation is computed from wind speed as:
\begin{align}
    P = \begin{cases}
        \text{max} \left( 0, \;\;\; P_r \left( v^3 - v^3_{min} \right) / \left( v^3_{max} - v^3_{min} \right) \right), \;\; & v < v_{min} \\
        P_r \left( v^3 - v^3_{min} \right) / \left( v^3_{max} - v^3_{min} \right), \;\; & v_{min} \leq v \leq v_{max} \\
        \text{min} \left( P_r, \;\;\; P_r \left( v^3 - v^3_{min} \right) / \left( v^3_{max} - v^3_{min} \right) \right), \;\; & v > v_{max}
    \end{cases}
\end{align}
where $v_{min}$ represents the minimum wind speed that can rotate wind turbine, and $v_{max}$ denotes wind speed at which maximum power generation limits is reached. $P_r$ is the wind power generation at $v_{max}$. In this work, $v_{min} = 1$~$ms^{-1}$, $v_{max} = 15$~$ms^{-1}$ and $P_r = 100$~$MW$.

\singlespacing
\begin{table}[!ht]
    \centering
    \caption{Case118 power grid}
    \label{tab:Case118_zone_info}
    \begin{tabular}{cccc}
        \toprule
         Zone & \# of dispatchable generators & \# of wind turbines & \# of loads \\
        \midrule
        I & 12 & 3 & 29 \\
        II & 15 & 7 & 38 \\
        III & 11 & 6 & 32 \\
        \bottomrule
    \end{tabular}
\end{table}

\begin{table}[!ht]
    \centering
    \caption{Distribution types and parameters for Case118.}
    \label{tab:Case118_PDF_parameters}
    \begin{tabular}{ccccccc}
    \toprule
        \multicolumn{2}{c}{Zone} & $\mu$ & $k$ & $\sigma$ & $a$ & $b$ \\
        \midrule
        \multirow{2}{*}{I} & TN & 50 & -- & 15 & 10 & 90 \\
        & WB & -- & 2 & 8 & -- & -- \\
        \multirow{2}{*}{II} & TN & 75 & -- & 20 & 25 & 125 \\
        & WB & -- & 1.8 & 8.2 & -- & -- \\
        \multirow{2}{*}{III} & TN & 100 & -- & 15 & 60 & 140 \\
        & WB & -- & 2.2 & 7.8 & -- & -- \\
    \bottomrule
    \end{tabular}
\end{table}

\doublespacing
\subsection{Case1354pegase} 

\begin{figure}[H]
    \centering
    \includegraphics[width=0.85\linewidth]{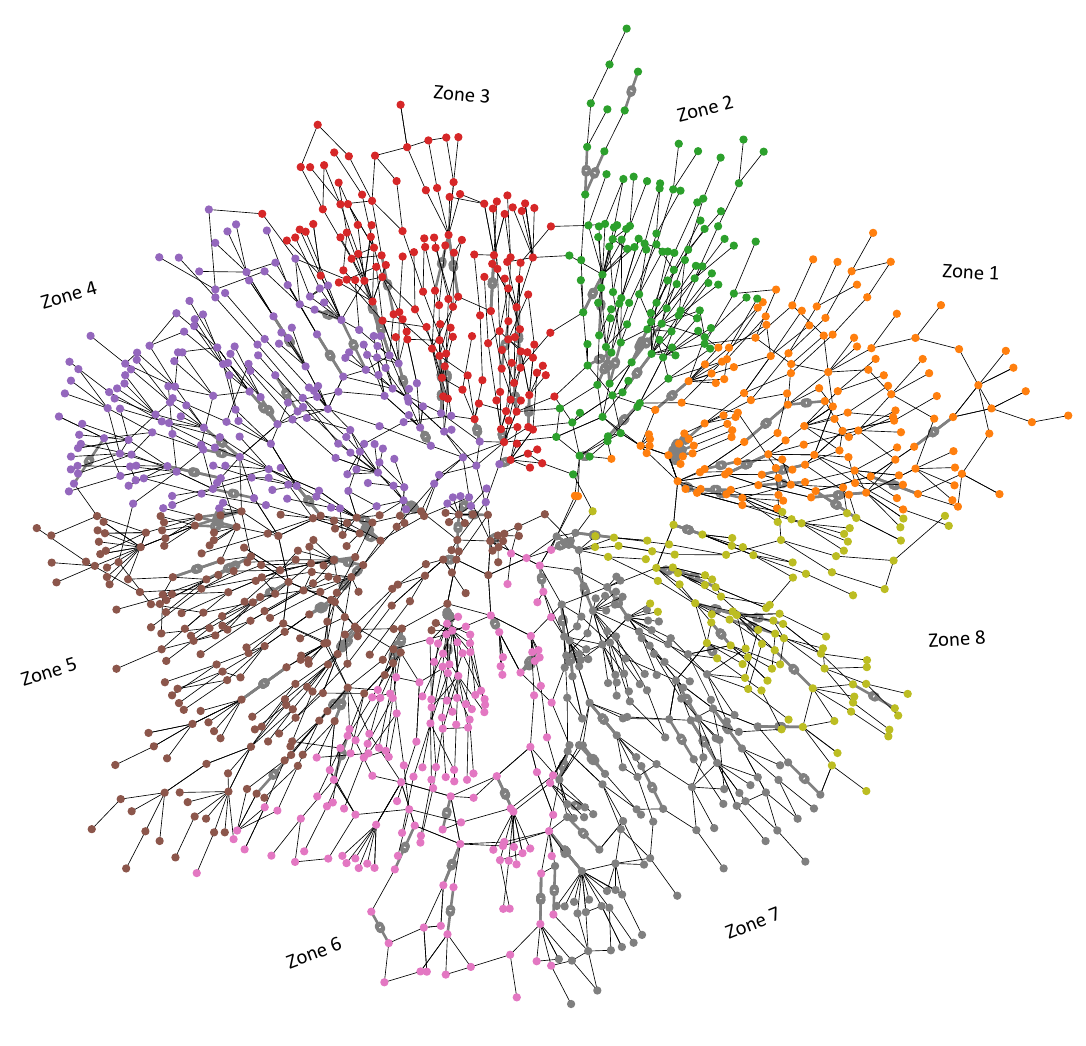}
    \caption{Case1354pesage power grid.}
    \label{fig:Case1354pesage_power_grid}
\end{figure}

The Case1354pegase power grid accurately represents the size and complexity of part of the European high voltage transmission network. The network contains 1354 buses (260 generators, 621 loads and 1991 branches). The data stems from the Pan European Grid Advanced Simulation and State Estimation (PEGASE) project, part of the 7th Framework Program of the European Union. Refer to~\citep{fliscounakis2013contingency} for more information. The grid is partitioned into 8 zones as shown in Fig.~\ref{fig:Case1354pesage_power_grid} and Table~\ref{tab:Case1354pegase_zone_info}. The probability distribution parameters are listed in Table~\ref{tab:Case1354_PDF_parameters}.

\singlespacing
\begin{table}[!ht]
    \centering
    \caption{Case1354pegase power grid}
    \label{tab:Case1354pegase_zone_info}
    \begin{tabular}{cccc}
        \toprule
        Zone & \# of dispatchable generators & \# of wind turbines & \# of loads \\
        \midrule
        I & 39 & 9 & 60 \\
        II & 32 & 8 & 48 \\
        III & 28 & 7 & 76 \\
        IV & 20 & 5 & 115 \\
        V & 24 & 5 & 114 \\
        VI & 11 & 2 & 94 \\
        VII & 34 & 8 & 62 \\
        VIII & 23 & 5 & 52 \\
        \bottomrule
    \end{tabular}
\end{table}

\begin{table}[!ht]
    \centering
    \caption{Distribution types and parameters for grid Case1354pegase.}
    \label{tab:Case1354_PDF_parameters}
    \begin{tabular}{ccccccc}
        \toprule
        \multicolumn{2}{c}{Zone} & $\mu$ & $k$ & $\sigma$ & $a$ & $b$ \\
        \midrule
        \multirow{2}{*}{I} & TN & 50 & -- & 15 & 10 & 90 \\
        & WB & -- & 2 & 8 & -- & -- \\
        \multirow{2}{*}{II} & TN & 75 & -- & 20 & 25 & 125 \\
        & WB & -- & 1.8 & 8.2 & -- & -- \\
        \multirow{2}{*}{III} & TN & 100 & -- & 15 & 60 & 140 \\
        & WB & -- & 2.2 & 7.8 & -- & -- \\
        \multirow{2}{*}{IV} & TN & 48 & -- & 12 & 9 & 87 \\
        & WB & -- & 1.9 & 8.1 & -- & -- \\
        \multirow{2}{*}{V} & TN & 81 & -- & 21 & 23 & 139 \\
        & WB & -- & 2.1 & 7.9 & -- & -- \\
        \multirow{2}{*}{VI} & TN & 98 & -- & 15 & 60 & 136 \\
        & WB & -- & 1.9 & 7.7 & -- & -- \\
        \multirow{2}{*}{VII} & TN & 52 & -- & 14 & 7 & 97 \\
        & WB & -- & 2.1 & 8 & -- & -- \\
        \multirow{2}{*}{VIII} & TN & 83 & -- & 20 & 21 & 145 \\
        & WB & -- & 2.0 & 8.3 & -- & -- \\
        \bottomrule
    \end{tabular}
\end{table}

\newpage
\doublespacing

\subsection{Case2848rte}

\begin{figure}[H]
    \centering
    \includegraphics[width=0.9\linewidth]{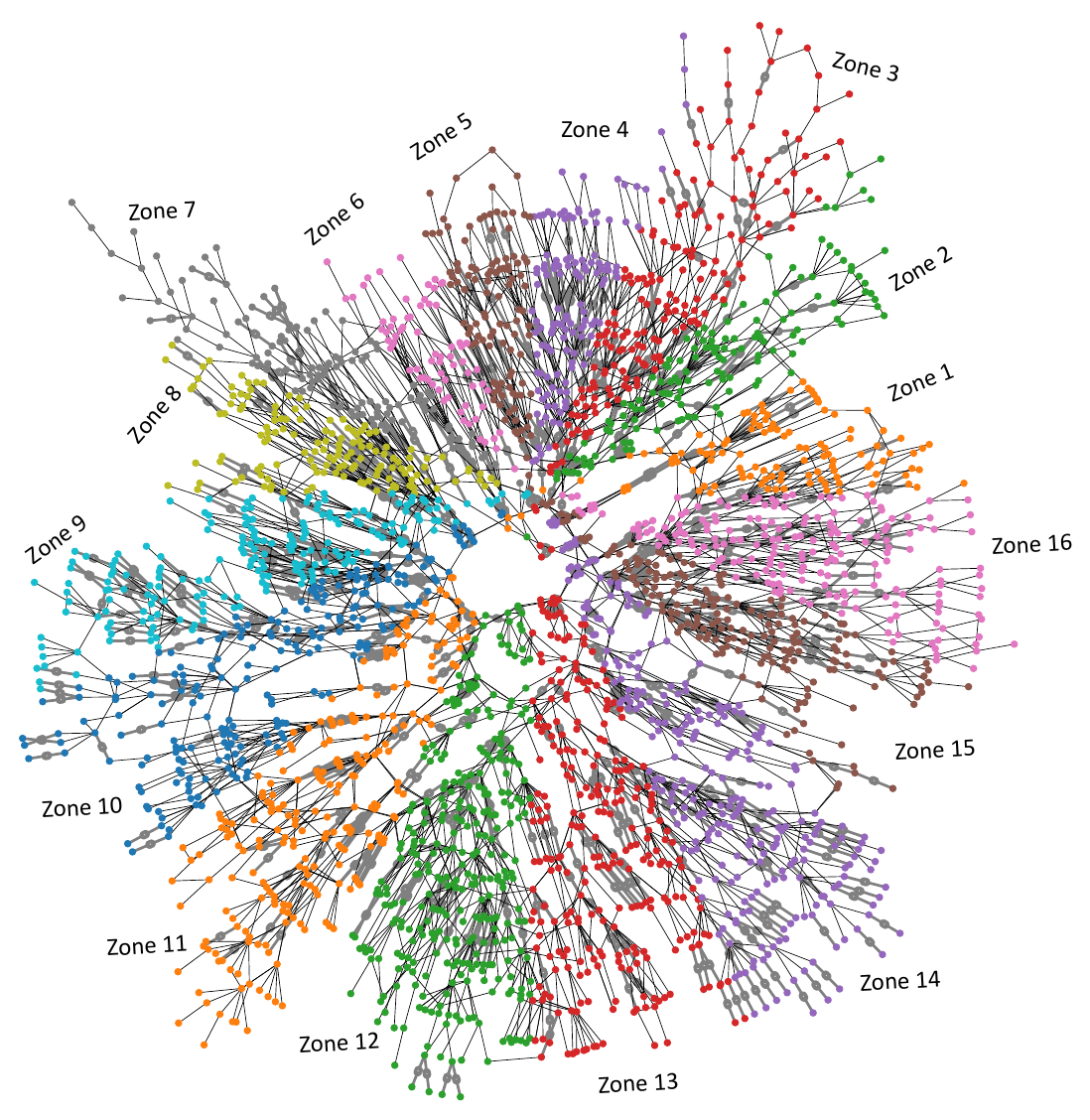}
    \caption{Case2848rte power grid.}
    \label{fig:Case2848rte_power_grid}
\end{figure}

Case2848rte power grid accurately represents the size and complexity of French (very high voltage) transmission network. The network contains 2848 buses (270 generators, 1388 loads and 3776 branches). Refer to~\citep{josz2016ac} for more details. The grid is partitioned into 16 zones (see Fig~\ref{fig:Case2848rte_power_grid} and Table~\ref{tab:Case2848pegase_zone_info}), and distribution parameters are shown in Table~\ref{tab:Case2848_PDF_parameters}.

\newpage

\singlespacing
\begin{table}[H]
    \centering
    \caption{Case28248rte power grid}
    \label{tab:Case2848pegase_zone_info}
    \begin{tabular}{cccc}
        \toprule
        Zone & \# of dispatchable generators & \# of wind turbines & \# of loads \\
        \midrule
        I & 23 & 5 & 43 \\
        II & 24 & 5 & 62 \\
        III & 22 & 5 & 78 \\
        IV & 8 & 2 & 71 \\
        V & 8 & 2 & 59 \\
        VI & 8 & 2 & 39 \\
        VII & 22 & 5 & 57 \\
        VIII & 16 & 3 & 69 \\
        IX & 21 & 5 & 104 \\
        X & 13 & 3 & 113 \\
        XI & 15 & 3 & 113 \\
        XII & 16 & 4 & 166 \\
        XIII & 20 & 5 & 149 \\
        XIV & 44 & 10 & 95 \\
        XV & 17 & 4 & 82 \\
        XVI & 24 & 6 & 88 \\
        \bottomrule
    \end{tabular}
\end{table}

\begin{table}[H]
    \centering
    %\small
    \caption{Distribution types and parameters for grid Case2848rte.}
    \label{tab:Case2848_PDF_parameters}
    \begin{tabular}{ccccccc}
        \toprule
        \multicolumn{2}{c}{Zone} & $\mu$ & $k$ & $\sigma$ & $a$ & $b$ \\
        \midrule
        \multirow{2}{*}{I} & TN & 50 & -- & 15 & 10 & 90 \\
        & WB & -- & 2 & 8 & -- & -- \\
        \multirow{2}{*}{II} & TN & 75 & -- & 20 & 25 & 125 \\
        & WB & -- & 1.8 & 8.2 & -- & -- \\
        \multirow{2}{*}{III} & TN & 100 & -- & 15 & 60 & 140 \\
        & WB & -- & 2.2 & 7.8 & -- & -- \\
        \multirow{2}{*}{IV} & TN & 48 & -- & 12 & 9 & 87 \\
        & WB & -- & 1.9 & 8.1 & -- & -- \\
        \multirow{2}{*}{V} & TN & 81 & -- & 21 & 23 & 139 \\
        & WB & -- & 2.1 & 7.9 & -- & -- \\
        \multirow{2}{*}{VI} & TN & 98 & -- & 15 & 60 & 136 \\
        & WB & -- & 1.9 & 7.7 & -- & -- \\
        \multirow{2}{*}{VII} & TN & 52 & -- & 14 & 7 & 97 \\
        & WB & -- & 2.1 & 8 & -- & -- \\
        \multirow{2}{*}{VIII} & TN & 83 & -- & 20 & 21 & 145 \\
        & WB & -- & 2.0 & 8.3 & -- & -- \\
        \multirow{2}{*}{IX} & TN & 45 & -- & 12 & 14 & 76 \\
        & WB & -- & 1.8 & 7.8 & -- & -- \\
        \multirow{2}{*}{X} & TN & 85 & -- & 18 & 25 & 145 \\
        & WB & -- & 1.8 & 8.1 & -- & -- \\
        \multirow{2}{*}{XI} & TN & 98 & -- & 16 & 55 & 141 \\
        & WB & -- & 8.1 & 12.5 & -- & -- \\
        \multirow{2}{*}{XII} & TN & 53 & -- & 13 & 9 & 97 \\
        & WB & -- & 2.2 & 8.5 & -- & -- \\
        \multirow{2}{*}{XIII} & TN & 84 & -- & 20 & 23 & 139 \\
        & WB & -- & 1.8 & 7.5 & -- & -- \\
        \multirow{2}{*}{XIV} & TN & 98 & -- & 15 & 55 & 141 \\
        & WB & -- & 2 & 7.7 & -- & -- \\
        \multirow{2}{*}{XV} & TN & 47 & -- & 12 & 8 & 86 \\
        & WB & -- & 2 & 8 & -- & -- \\
        \multirow{2}{*}{XVI} & TN & 85 & -- & 20 & 21 & 149 \\
        & WB & -- & 2.5 & 8.3 & -- & -- \\
        \bottomrule
    \end{tabular}
\end{table}

\newpage

\section{GNN Model Training}
\label{sec:GNN_details}

% Add "S" in section number
% \renewcommand{\thesection}{S\arabic{section}}
% % Add "S"
% \renewcommand{\thefigure}{S-\arabic{figure}}   
\setcounter{figure}{0}   
\setcounter{table}{0}

\subsection{Number of parameters in each layer}
\label{sec:number_layer_parameter}

\begin{table}[H]
    \centering
    \caption{\# Neurons in each layer of GNN model}
    \label{tab:GNN_details}
    \begin{tabular}{cccc}
        \toprule
        \multirow{2}{*}{Layer} & & Dimension (Input channel, Output channel) & \\
        \cmidrule{2-4}
        & System-level & Zone-level & Branch-level \\
        \midrule
        Encoder-1 & $(D_I, \, 2D_I)$ & $(D_I, \, 2D_I)$ & $(D_I, \, 2D_I)$ \\
        Encoder-2 & $(2D_I, \, D_H)$ & $(2D_I, \, D_H)$ & $(2D_I, \, D'_H)$ \\
        GraphSAGE-1 & $(D_H, \, D_H)$ & $(D_H, \, D_H)$ & $(D'_H, \, D'_H)$ \\
        GraphSAGE-2 & $(D_H, \, D_H)$ & $(D_H, \, D_H)$ & $(D'_H, \, D'_H)$ \\
        Decoder-1 & $(D_H, \, 2T)$ & $(D_H, \, 2T)$ & $(D'_H, \, 2T)$ \\
        Decoder-2 & $(2T, \, T)$ & $(2T, \, T)$ & $(2T, \, T)$ \\
        Mean Pooling & $(T, \, T)$ & $(T, \, T)$ & N/A \\
        \bottomrule
    \end{tabular}
    \caption*{NOTE: $D_I = N_f + 2T$ where $N_f=8$ is the number of selected node features and $T=12$ is the number of time steps; $D_H=2D_I$, and $D_H' = 4D_I$; the prediction branch-level GNN model is concerted into branch flow using Power Transmission Distribution Matrix.}
\end{table}

\subsection{Training loss}
\label{sec:training_loss}

\begin{figure}[!ht]
    \centering
    \includegraphics[width=0.8\linewidth]{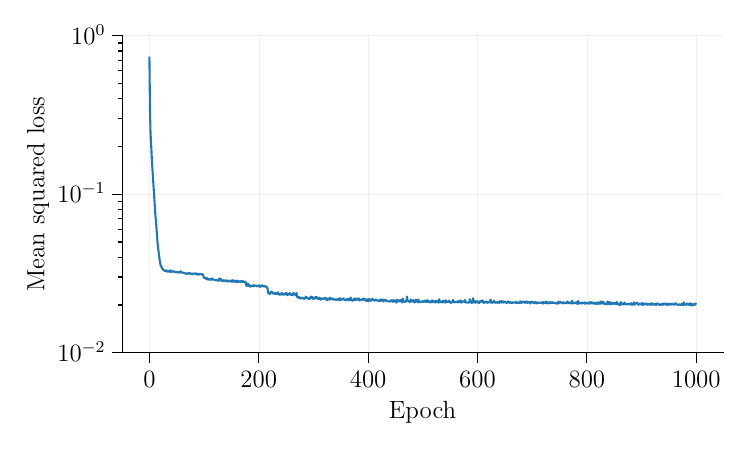}
    \caption{Training loss \textit{vs} epochs}
    \label{fig:epoch_loss}
\end{figure}

\newpage

\begin{landscape}

\section{Maximum Prediction Error}
\label{sec:max_prediction_error}

% Add "S" in section number
% \renewcommand{\thesection}{S\arabic{section}}
% % Add "S"
% \renewcommand{\thefigure}{S-\arabic{figure}}   
\setcounter{figure}{0}   
\setcounter{table}{0}

\begin{table*}[!ht]
    \centering
    \caption{Maximum relative error (\%) of dispatch prediction}
    \label{tab:max_dispatch_error}
    \begin{tabular}{ccccccccccccccccc}
        \toprule
        Grid & I & II & III & IV & V & VI & VII & VIII & IX & X & XI & XII & XIII & XIV & XV & XVI \\
        \midrule
        Case118 & 8.86 & 5.10 & 8.17 \\
        Case1354pegase & 8.61 & 6.46 & 9.59 & 8.57 & 7.71 & 5.71 & 6.87 & 8.37 \\
        Case2848rte & 9.54 & 6.60 & 5.45 & 6.50 & 5.57 & 9.14 & 5.23 & 8.13 & 7.74 & 9.10 & 5.99 & 9.28 & 6.76 & 8.77 & 6.48 & 9.42 \\
        \bottomrule
    \end{tabular}
\end{table*}

\begin{table*}[!ht]
    \centering
    \caption{Maximum relative error (\%) of load shedding prediction}
    \label{tab:max_shedding_error}
    \begin{tabular}{ccccccccccccccccc}
        \toprule
        Grid & I & II & III & IV & V & VI & VII & VIII & IX & X & XI & XII & XIII & XIV & XV & XVI \\
        \midrule
        Case118 & 8.86 & 5.10 & 8.17 \\
        Case1354pegase & 8.61 & 6.46 & 9.59 & 8.57 & 7.71 & 5.71 & 6.87 & 8.37 \\
        Case2848rte & 9.54 & 6.60 & 5.45 & 6.50 & 5.57 & 9.14 & 5.23 & 8.13 & 7.74 & 9.10 & 5.99 & 9.28 & 6.76 & 8.77 & 6.48 & 9.42 \\
        \bottomrule
    \end{tabular}
\end{table*}

\begin{table*}[!ht]
    \centering
    \caption{Maximum relative error (\%) of branch flow prediction}
    \label{tab:max_branch_flow_error}
    \begin{tabular}{ccccccccccccccccc}
        \toprule
        Grid  & 1 & 2 & 3 & 4 & 5 & 6 & 7 & 8 & 9 & 10 & 11 & 12 & 13 & 14 & 15 & 16 \\
        \midrule
        Case118 & 14.18 & 15.92 & 8.08 & 12.00 \\
        Case1354pegase & 10.14 & 6.53 & 6.43 & 12.34 & 16.43 & 13.80 & 11.39 & 8.35 \\
        Case2848rte & 16.51 & 8.46 & 12.42 & 13.18 & 12.78 & 17.82 & 8.45 & 6.20 & 17.88 & 16.40 & 12.39 & 11.84 & 7.63 & 10.74 & 7.33 & 16.02 \\
        \bottomrule
    \end{tabular}
\end{table*}
    
\end{landscape}
\end{document}